\documentclass[12pt]{iopart}
\usepackage{graphicx,color}
\usepackage{epstopdf}
\usepackage{bm}
\usepackage{subfig}
\usepackage{eucal}
\usepackage{mathrsfs}

\usepackage{nicefrac} 
\usepackage{xfrac}
\usepackage{hyperref}
\usepackage{cancel}
\usepackage[utf8]{inputenc}
\usepackage{float}
\usepackage{array}
\newcolumntype{P}[1]{>{\centering\arraybackslash}p{#1}}

\usepackage{iopams}  
\begin{document}

\title[Decreased resilience in power grids under dynamically induced vulnerabilities]{Decreased resilience in power grids under dynamically induced vulnerabilities}

\author{C. C. Galindo-González$^1$, D. Angulo-Garcia$^2$ and G.~Osorio$^1$}

\address{$^1$ Grupo de Percepción y Control Inteligente (PCI). Departamento de Ingeniería Eléctrica, Electrónica y Computación. Universidad Nacional de Colombia - Sede Manizales. Kilómetro 9, vía al aeropuerto La Nubia. Manizales, Colombia.}
\address{$^2$ Grupo de Modelado Computacional - Dinámica y Complejidad de Sistemas. Instituto de Matemáticas Aplicadas. Universidad de Cartagena. Carrera 6 \# 36 - 100. 130001. Cartagena de Indias, Colombia}
\ead{\mailto{ccgalindog@unal.edu.co}, \mailto{dangulog@unicartagena.edu.co}, \mailto{gaosoriol@unal.edu.co} }

\vspace{10pt}
\begin{indented}
\item[]July 2020
\end{indented}

\begin{abstract}
In this paper, a methodology inspired on bond and site percolation methods is applied to the estimation of the resilience against failures in power grids. Our approach includes vulnerability measures with both dynamical and structural foundations as an attempt to find more insights about the relationships between topology and dynamics in the second-order Kuramoto model on complex networks. As test cases for numerical simulations, we use the real-world topology of the Colombian power transmission system, as well as randomly generated networks with spatial embedding. It is observed that, by focusing the attacks on those dynamical vulnerabilities, the power grid becomes, in general, more prone to reach a state of total blackout, which in the case of node removal procedures it is conditioned by the homogeneity of power distribution in the network. 
\end{abstract}

\vspace{2pc}
\noindent{\it Keywords}: Kuramoto model, power grid, basin stability, resilience, stability, percolation.
%
%
%
%

\section{Introduction}

Emergence of synchronization in systems of agents, coupled through local interactions, is a widely studied phenomenon with multiple applications in physics, biology and social sciences \cite{PIK01, ARENAS200893, okz07}. In particular, the optimal functioning of a power grid, which represents one of the most complex interacting systems in engineering, highly depends on its ability to maintain a synchronous operation over time despite external disturbances. Losing that synchrony, even locally, may lead to cascading failures and complete blackout of the network \cite{dorfler2013synchronization, Motter2013, Sakaguchi2012}. Two main concerns have motivated the discussion around dynamical analysis and design of power grids in the last years, those are: the strong economical and social impact that a power outage could cause in the highly electricity-dependant modern society \cite{Buldyrev2010}, and the slow transition to renewable energy sources that is being promoted all around the world \cite{brockway_renew, Xi2018}, since it is known that renewable and small power producers have negligible inertia, which imposes instability risks in the power grids \cite{Tamrakar2018}.     
Multiple models have been proposed for the analysis of synchronization in power grids, and they differ mainly in the way loads, power generators and transmission lines are modeled \cite{Nishikawa2015}. Probably the most common model is the one called synchronous motor, where both, generators and consumers are represented by second-order synchronous machines \cite{Filatrella2008}. This model is essentially equivalent to the celebrated Kuramoto model when inertia terms are considered \cite{kuramoto1975, acebron_bonilla, Dorfler2010a}. Self-synchronization behavior in this model occurs when the interactions between connected oscillators are sufficiently strong to overcome the dissimilarity in the ensemble \cite{ dorfler2013synchronization, rohden2014NJP, rohden2012PRL, Olmi2014}.

Recently, the scientific community has put a lot of effort in trying to determine the topological or structural features of those interactions that can enhance or undermine non-linear stability of a power grid, that is, its capability to reject finite-size disturbances, a matter of paramount importance in the design of the future smart grids. This has been approached, for instance, by the means of energy barrier functions \cite{Xi2017} and the basin stability concept \cite{Menck2013, Menck2014, Nitzbon2017, Mitra2017b, Mitra2017a, Wolff_Lind_2018}. Some interesting findings that deserve to be mentioned, as they provide great insights about the relationship between topology and dynamical stability include: the poor basin stability usually detected on dead-tree arrangements \cite{Menck2013}, the strong stability found on triangle-shaped motifs \cite{Schultz2014_Detour}, the enhanced non-linear stability achieved by increasing global redundancy in the connections \cite{Plietzsch2016} or by adding small cyclic motifs \cite{Xi2017} and the lower basin stability exhibited in general by high-power generator nodes in \cite{Montanari2020}.
Other studies have focused on estimating the resilience of the complex network against cascading failures  \cite{Simonsen2008, Sakaguchi2012, Rohden2016}, the robustness in the transient behavior after localized disturbances \cite{Schafer2018} and their diffusion through the interconnected system \cite{Tamrakar2018}.

In the present work, a percolation inspired method is presented in order to evaluate the resilience of a power grid against random and focalized disturbances. By defining a vulnerability measure in terms of the dynamical properties of both, nodes and edges, it is found that, attacks that focus on the weakest components of the network, can easily provoke cascading failures that lead to total blackout faster than random attacks in some specific cases. Note that, in the following, the term total blackout refers to the situation when no operational composition of at least one generator and one consumer exists in the network. Test cases chosen for this study include randomly generated networks with spatial embedding that emulate properties of real-world power grids, as well as the particular case of the Colombian power transmission network. It is also found that, by attacking the most dynamically-loaded transmission lines in the graph, the power system reaches blackout faster than any other bond-percolation criterion, while, for the site-percolation methods, it is produced by attacking the most dynamically vulnerable nodes, but only if the distribution of power generation and consumption is not homogeneous, which often occurs in real-world architectures.

A similar work can be found in \cite{MachucaMoreno2017}, where the authors use centrality measures, namely, the node degree, clustering coefficient and betweenness centrality, to estimate the vulnerability of a node and then proceed to remove weakest nodes to observe the evolution of the giant component in the network as a function of the number of remaining nodes in the whole graph. This structural approach to quantify the node vulnerability will be also used in this work to compare results of removal methods based on dynamical measurements. Other relevant works have to be mentioned; for instance, based on the statistical physics description of networks, some basic percolation properties for different networks have been analysed using generating functions \cite{Fu2017}, the lifetime and reliability of networks has been estimated by setting some probability distributions for node failures \cite{Li2015} and the propagation of cascading failures through the power transmission or communication architecture of smart grids has also been modeled \cite{Huang2013}.

The rest of this paper is organized as follows: Section \ref{sec:mod_methods} introduces the dynamical model of a power grid and the structure of the specific test cases, as well as some theoretical tools that will be applied in further analysis of synchronization and vulnerability measurements. Section \ref{sec:resilresul} presents an algorithm for resilience assessment of power grids and results regarding line and node removal procedures over synthetic power grids and the Colombian transmission network. Finally, some concluding remarks are discussed in section \ref{sec:conclus}.

\section{\label{sec:mod_methods}Model and methods}

In this paper, the framework of the synchronous motor representation is used to model the general structure and dynamics of power grids \cite{Nishikawa2015, Filatrella2008}, since it has been employed constantly and successfully in the past years, as mentioned in the previous section. This approach is described in the following.

\subsection{\label{sec:powerGrid} Power grid model}
As a simplified model of a real-world power grid, consider a system
of $N$ interacting synchronous machines arranged in a connected graph $G{(\vartheta,\varepsilon)}$, with a set of nodes $\vartheta=\{1,2,...,N\}$ and a set of edges
$\varepsilon\subset\vartheta\times\vartheta$, such that the amount of edges is $|\varepsilon| = M$. Nodes can be labeled as generator machines (supply energy to the grid) or consumer machines (demand energy from the grid), thus $\vartheta=\vartheta_g \cup \vartheta_c$, being $\vartheta_g$ the set of generators and $\vartheta_c$ the set of consumers.

Let the dynamical state of each machine be represented by its phase angle $\phi_i$ and its phase velocity $\dot \phi_i : = {\sfrac{d \phi_i}{dt} }$, $i\in\vartheta$. Under an appropriate operation of the power grid, it is expected that every machine will be rotating at some reference frequency $\Omega=2\pi f$ (where by convention $f$ is either $50$ Hz or $60$ Hz), so let $\theta_{i (t)} = \phi_{i (t)} - \Omega t$ be the phase deviation of the machine $i$ with respect to the reference angle $\Omega t$, then the dynamical behavior of $\theta_{i}$ and its velocity $\dot{\theta}_{i}$ can be described by the well-known second-order Kuramoto model for coupled oscillators given by \cite{Filatrella2008, rohden2014NJP}:
\begin{equation}
\label{eq:kuramoto}
\ddot{\theta}_{i (t)}=P_{i}-\alpha_{i}\dot{\theta}_{i (t)}+K\sum\limits _{j}^{N}{a_{ij}\sin(\theta_{j (t)}-\theta_{i (t)})}
\end{equation}
where $P_{i}$ is equal to the injected (drained) power at the $i$-th node up to a scaling factor, and it is positive (negative) for generators (consumers);  $\alpha_{i}$ is related to the damping coefficient of the $i$-th machine, $a_{ij}$ are the elements of the adjacency matrix $A$, thus, $a_{ij}=1$ if nodes $i$ and $j$ are connected (that is, $(i,j) \in \varepsilon$) and $a_{ij}=0$ otherwise. The constant $K$ is the maximum power transfer capacity between any pair of connected nodes, which depends on the impedance of transmission lines connecting them. This equation is derived when the ohmic losses on transmission lines are neglected, an equal impedance level is assumed for every transmission line in the network and the deviations from the phase velocity at any node as compared to the reference $\Omega$ are small, that is $|\dot{\theta}_{i}|\ll\Omega$, $\forall i \in \vartheta$ \cite{Filatrella2008, rohden2014NJP}.
\begin{figure*}
\includegraphics[width=1.0\linewidth]{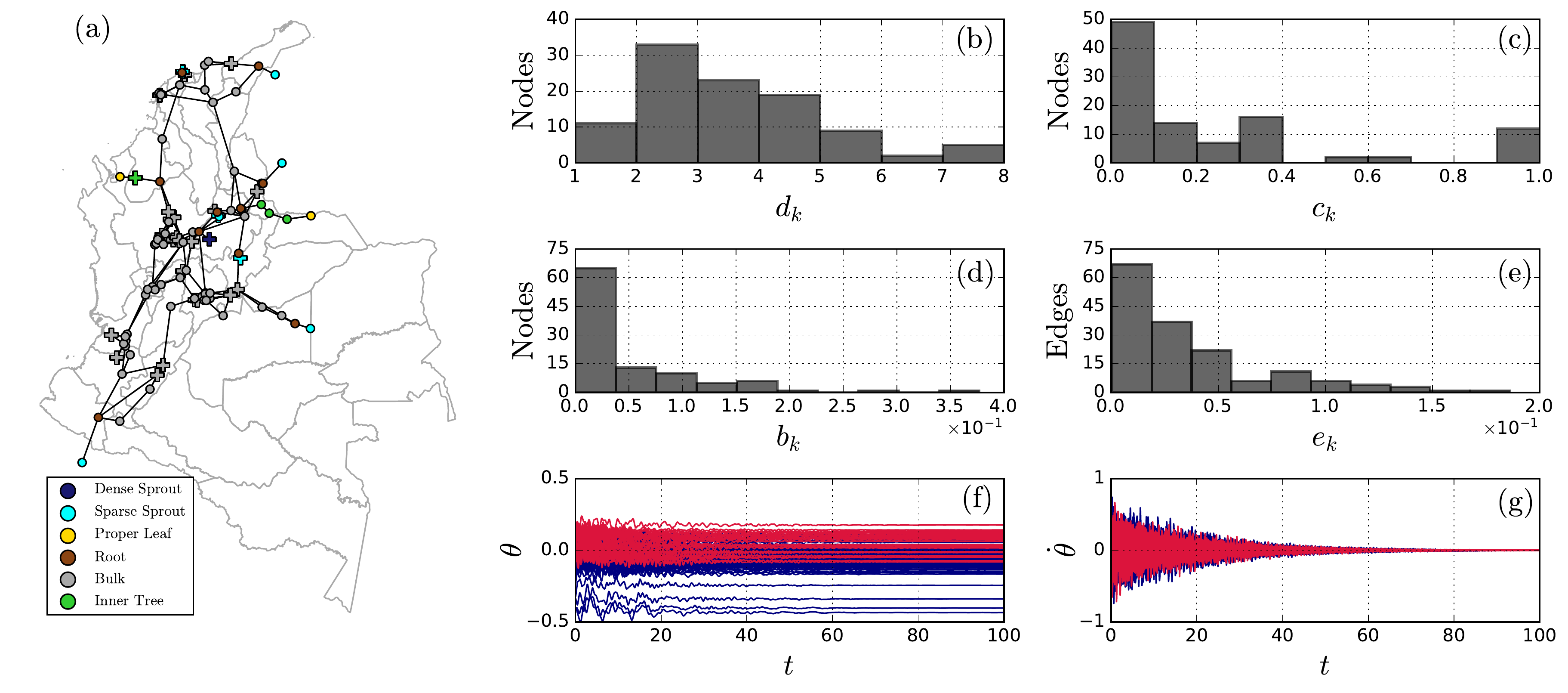}
\caption{\text{Colombian Power Grid:} (a) Topology of the network with $N=102$ and $M = 158$. Generating nodes $N_g = 28$ (cross symbols) and $N_c = 74$ consumer nodes (circle symbols). (b)-(e) Distribution of the node degree $d_k$, clustering coefficient $c_k$, node betweenness centrality $b_k$ and edge betweenness centrality $e_k$. (f)-(g) Time trace of the phases and phase velocities of each oscillator when the grid converges to a frequency synchronized state (red lines: generators, blue lines: consumers).
\label{fig:ColombianPowerGrid}}
\end{figure*}

The dynamical system described by equation \eref{eq:kuramoto} is said to be in a phase synchronized state if $\theta_{i (t)}=\theta_{j (t)}$, $\text{\ensuremath{\forall}}i,j\in\vartheta$, and in a frequency synchronized state if $\dot{\theta}_{i (t)}=\dot{\theta}_{j (t)}= \omega_s$, $\text{\ensuremath{\forall}}i,j\in\vartheta$. Since we are interested in the dynamics around the synchronization frequency $\Omega$, it can be assumed  without loss of generality that $\omega_s = 0$, which amounts to consider the dynamics in the co-rotating reference frame $\Omega t$ (see \cite{dorfler2013synchronization}).
In the following, the topology of $G{(\vartheta,\varepsilon)}$ is taken from the Colombian National Transmission System \cite{upme}, a power grid composed of 158 edges and 102 nodes (28 of them are generators and 74 are consumers). Power supplied by generators is set to $P_k=1.0$, $\forall k \in \vartheta_g$, and the power drained by consumers is uniformly distributed such that the power balance condition in the network is fulfilled, that is:
\begin{equation}
\sum_{i=1}^{N}P_{i}=0.
\label{eq:powerbalance}
\end{equation}

This condition is a requirement in order to allow the existence of a locally-stable equilibrium point in the system \cite{rohden2014NJP}. Finally, the damping and maximum power transfer are fixed to $\alpha_{i} = 0.1$ and $K=12.0$ respectively. Figure \ref{fig:ColombianPowerGrid}(a) shows the topology of the network with the tree-like node classification introduced in \cite{Nitzbon2017}. Panels (b)-(e) present distributions for some classic centrality measures of the graph: node degree, clustering coefficient \cite{Saramaki2007}, node betweenness and edge betweenness centrality \cite{ulrikb}. Panels (f) and (g) display the behavior of $\theta_{(t)}$ and $\dot \theta _{(t)}$ for every oscillator when the system converges from an unsynchronized state to a frequency synchronized regime given the parameters above.  
Aiming for a generalization of the results, random test cases were also used and compared to the Colombian test case just described; to do that, synthetic power grids were generated with the algorithm proposed in \cite{Schultz2014}, which is specially useful as it creates spatially embedded networks that emulate connectivity distributions of real-world power grids. Synthetic networks were constructed by initializing a tree-like base architecture and then following a growth procedure conditioned by the optimization of a redundancy-cost function that depends on both, geographic distance and number of paths separating each pair of nodes in the graph. For a detailed description of the algorithm, refer to \ref{random_grids}. The parameters used for the growth model algorithm were chosen to match those used on \cite{Nitzbon2017}, as presented in table \ref{tab:synthetic_grids_params}. Note that the final size of the network is fixed to $N = 102$ so that it matches the Colombian power grid size, nevertheless, geographic positions for nodes are chosen uniformly at random and an equal amount of generators and consumers is placed to reduce heterogeneity in the samples (thus, $P_k=1.0$, $\forall k \in \vartheta_g$ and $P_j=-1.0$, $\forall j \in \vartheta_c$).

\begin{table}[h]
\centering
\caption{\label{tab:synthetic_grids_params}Parameters for the construction of synthetic power grids. }
\begin{tabular}{|P{12cm}|P{2cm}|} 
\hline
\textbf{Parameter} & \textbf{Value}  \\ 
\hline
$N_0$: An initial amount of nodes, $N_0 \geq 1$. & 1\\
\hline
$N$: Final amount of nodes that the network will have $N > N_0$. & 102\\
\hline
$y = \{ y_1, y_2, ..., y_{N} \}$: Geographical location of nodes. & $y_k \sim U_{[0,1]}$\\
\hline
$p$: Probability of constructing additional redundancy links attached to new nodes ($0 \leq p \leq 1$). & $\nicefrac{1}{5}$\\
\hline
$q$: Probability of constructing additional redundancy links between existing nodes ($0 \leq q \leq 1$). & $\nicefrac{3}{10}$\\
\hline
$s$: Probability of splitting an existing line ($0 \leq s \leq 1$). & $\nicefrac{1}{10}$\\
\hline
$u$: Cost-vs-redundancy trade-off parameter for equation \ref{eq:optimcost}. & $\nicefrac{1}{3}$\\
\hline
\end{tabular}
\end{table}

\subsection{Synchronization measures }
A helpful indicator that quantifies the degree of synchronization is the complex-valued order parameter $r$ defined as \cite{Kuramoto2003}:
\begin{equation}
\label{eq:order_parameter}
r e^{i \Psi_{(t)}}=\frac{1}{N}\sum_{j}^{N}e^{i\theta_{j (t)}}
\end{equation}

Here, $\Psi_{(t)}$ is the average phase of the oscillators at time $t$. If all phases $\theta_{j}$ are identical, the magnitude of the order parameter is $r_{(t)}=1$. Conversely, if they are equally distributed around the unit circle $r_{(t)}=0$ and the system is said to be desynchronized. Intermediate values of $r$ correspond to partially synchronized states. Additionally, in the case of power grids in which equation \eref{eq:powerbalance} holds, when the system is synchronized the average phase is $\Psi_{(t)} = 0$, and the real part of the order parameter ${\rm I\!R} [r] = 1$, while it oscillates around zero otherwise. While the dependence of the order parameter with time is useful in transient analysis, we will focus on the steady state behavior, so we can define the average order parameter in steady state as \cite{rohden2014NJP}:

\begin{equation}
r_{\infty}=\lim_{t_{1}\rightarrow\infty}\left[\lim_{t_{2}\rightarrow\infty}\left(\frac{1}{t_{2}}\int_{t_{1}}^{t_{1}+t_{2}}r_{(t)}dt\right)\right]
\end{equation}

Another measure of frequency synchronization in power grids is the squared average rotational speed $v_{(t)}^{2}$ defined as \cite{rohden2014NJP}:
\begin{equation}
v_{(t)}^2=\frac{1}{N}\sum_{j}^{N}\dot{\theta}_{j (t)}^{2}\label{eq:v_squared}
\end{equation}
and the associated steady state value of the speed $v_{\infty}$:
\begin{equation}
v_{\infty}=\sqrt{\lim_{t_{1}\rightarrow\infty}\left[\lim_{t_{2}\rightarrow\infty}\left(\frac{1}{t_{2}}\int_{t_{1}}^{t_{1}+t_{2}}v_{(t)}^{2}dt\right)\right]}.
\end{equation}

\subsection{Edge vulnerability}

Following the results in \cite{dorfler2013synchronization}, it can be proven that, by applying a small-angle approximation, the phases at the frequency synchronized equilibrium point, which are denoted by $\theta^* \in {\rm I\!R}^N$, can be estimated as:

\begin{equation}
\theta^* \approx \frac{1}{K}L^\dagger P    
\end{equation}

where $L^{\dagger}$ is the pseudo-inverse of the Laplacian matrix, which is computed through the adjacency matrix as $L = diag( {\sum_{j=1}^n a_{ij}} ) - A$. Let us now define the oriented incidence matrix $B = \{ b_{ij} \} \in \mathbb{R}^{N \times M}$, with:
\begin{equation}
b_{ij} = \cases{
1 & $\textrm{if} ~ \textrm{node} ~ i ~ \textrm{incides} ~ \textrm{in} ~ \textrm{edge} ~ j$\\
-1 & $\textrm{if} ~ \textrm{edge} ~ j ~ \textrm{incides} ~ \textrm{in} ~ \textrm{node} ~ i$\\
0 & \textrm{otherwise}
}
\label{eq:inc_matrix}
\end{equation}

Note that although $G$ is an undirected graph, a direction for each edge can be assumed without loss of generality for the following analysis, allowing for this definition of $B$. Under these circumstances, the steady state phase difference between any pair of connected nodes $(i, j) \in \varepsilon$ can be approximated by:
\begin{equation}
\label{eq:phase_diff}
\Delta_\theta \approx \frac{1}{K}B^TL^\dagger P
\end{equation}
with $\Delta_\theta \in \mathbb{R}^M$. Then, the system \eref{eq:kuramoto} is guaranteed to have a unique and stable solution with synchronized frequencies and cohesive phases if the next sufficient condition for $\Delta_\theta$ holds:
\begin{equation}
\label{eq:dorfler_sync_condition}
|| \Delta_\theta ||_{\infty} < \sin (\gamma)
\end{equation}
with $0 \leq \gamma < \pi / 2$. In the limit $\gamma \to \pi/2$, equation \eref{eq:dorfler_sync_condition} simply reduces to $|| \Delta_\theta ||_{\infty} < 1$. In other words, equation \eref{eq:dorfler_sync_condition} states that, in order to achieve synchronization, the largest difference between the phases of connected pairs in the network has to be smaller than $\nicefrac{\pi}{2}$ \cite{dorfler2013synchronization}. This rather simple equation, readily connects topological and dynamical features of the network. While condition \eref{eq:dorfler_sync_condition} is only sufficient, it is a rapid way to assess synchronization without relying on the computationally expensive time evolution of a large number of phase oscillators. Moreover, equation \eref{eq:dorfler_sync_condition} provides an approximation to the critical value $K_c$ of the coupling strength when substituting from \eref{eq:phase_diff}:
\begin{equation}
\label{eq:Dorfler_critical}
K_c \approx || B^TL^\dagger P ||_{\infty}
\end{equation}

\begin{figure}
\includegraphics[width=0.50\linewidth]{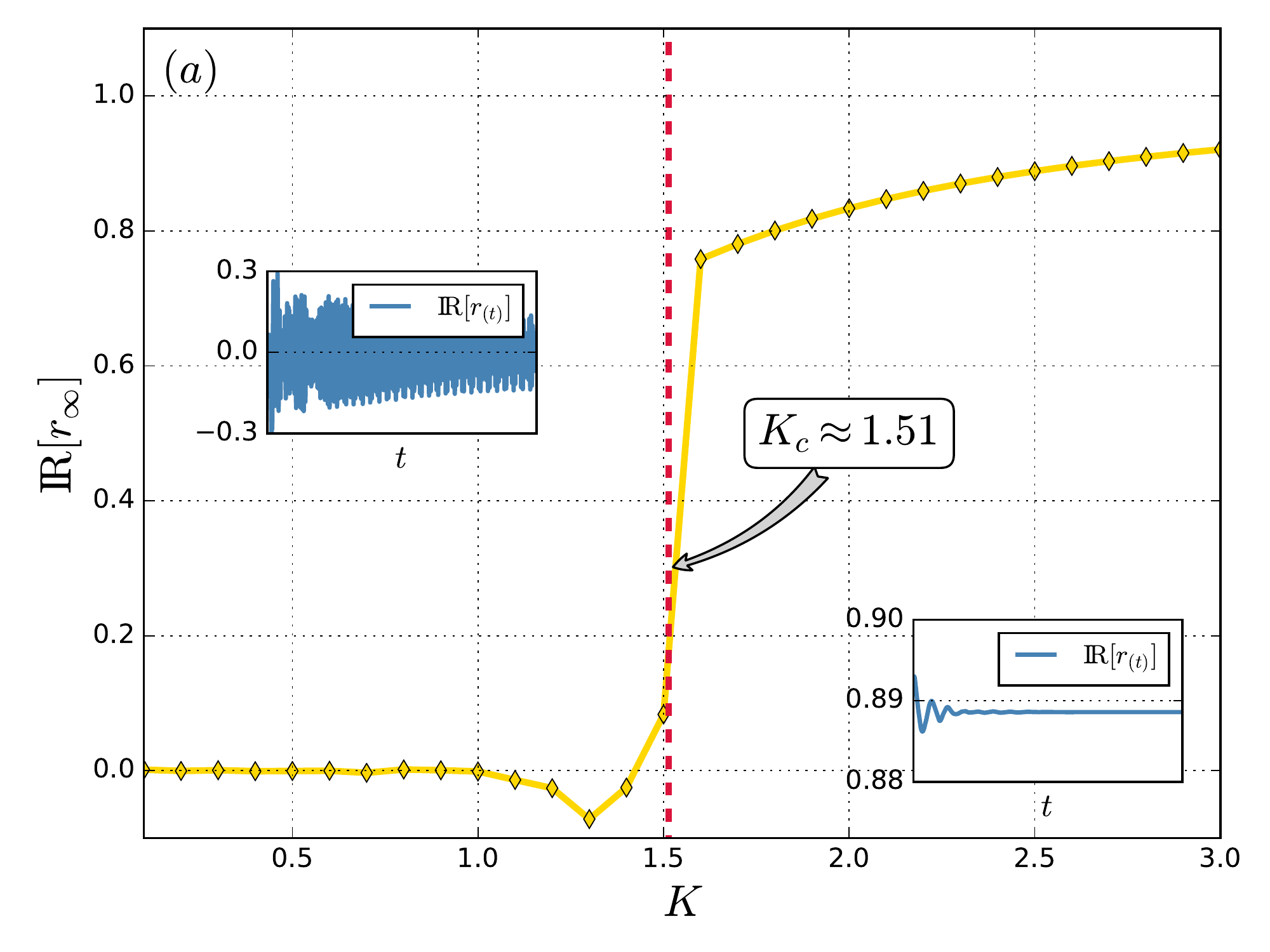}
\includegraphics[width=0.50\linewidth]{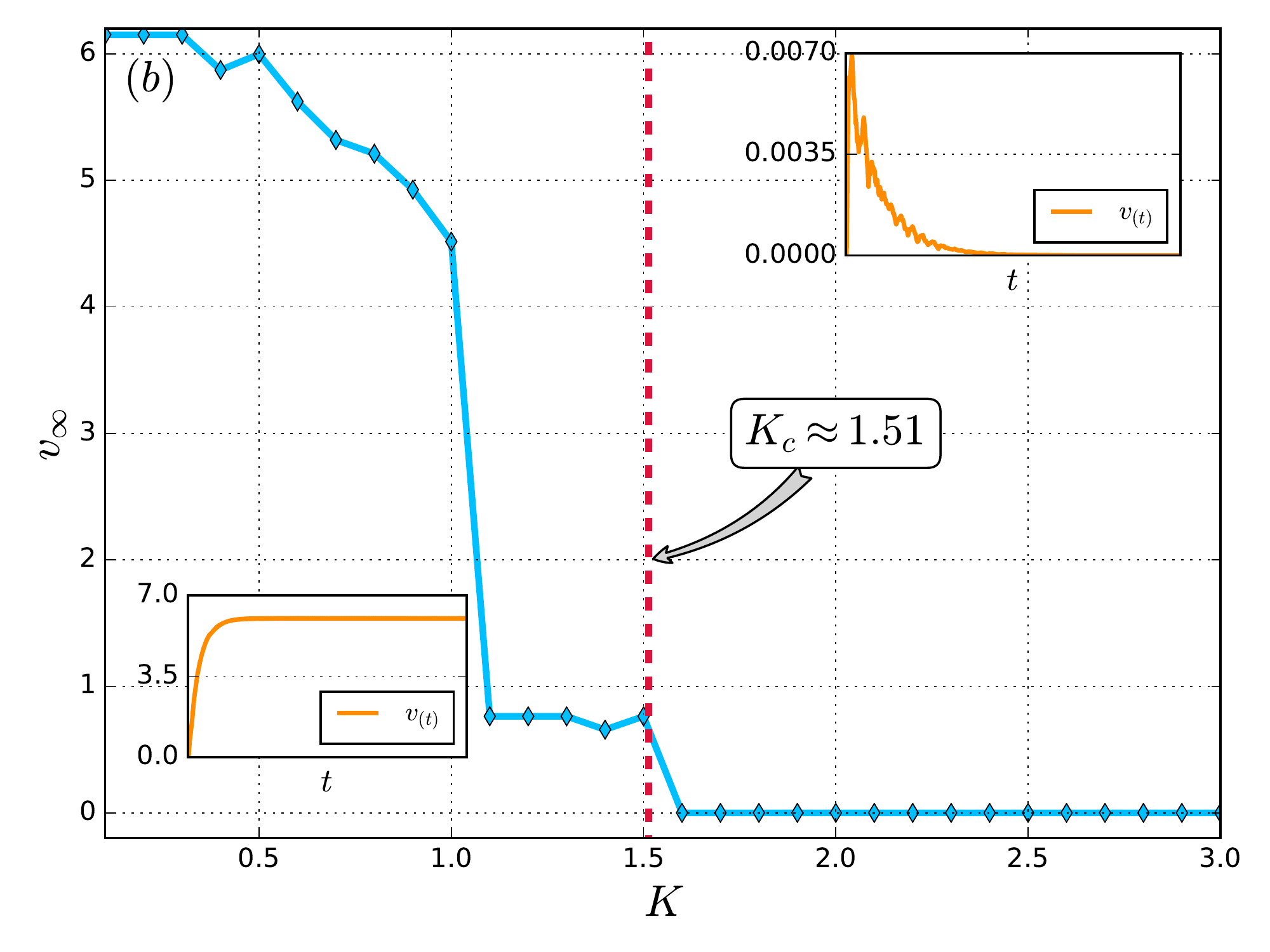}
\caption{\text{ Synchronization for the Colombian power grid:} (a) Average steady state value of the real part of the order parameter as a function of the coupling strength $K$. Left (right) inset shows the time trace of ${\rm I\!R} [r_{(t)}]$ before (after) critical coupling. (b) Average rotational speed as a function of $K$. As in (a), insets show time traces before and after critical coupling. In both cases, vertical dashed line depicts the value of $K_c$ calculated with equation \eref{eq:Dorfler_critical}.
\label{fig:synchronization_measures}}
\end{figure}

In figure \ref{fig:synchronization_measures}(a), it is shown the behavior of the steady state order parameter ${\rm I\!R}[r_{\infty}]$ as a function of the coupling strength $K$. As seen in the figure and the insets, the real part of the order parameter oscillates around zero when the coupling strength is small, and therefore its average value is zero. At a critical coupling strength $K_c \approx 1.51$, the system achieves a state of phase cohesiveness characterized by ${\rm I\!R}[r_{\infty}] \approx 1$ indicating a synchronized regime. The same transition can be observed by looking at the average velocity in figure \ref{fig:synchronization_measures}(b). In particular, it can be seen that before the synchronization transition, a finite and different than zero, steady-state value for the angular velocity is obtained, indicating that the phases of the oscillators increase constantly and no equilibrium is reached. After the transition, $v_{(t)}$ abruptly decreases to zero, meaning that the phases of the oscillators remain in a steady state.

In both figures, it is possible to see as well the accuracy of the sufficient condition \eref{eq:Dorfler_critical}, where the critical value is indicated by the vertical dashed line, showing a good agreement with the numerical value of the transition. This evidence supports the use of condition \eref{eq:dorfler_sync_condition} as a synchronization requisite in the following analysis. Note that the chosen value of $K=12.0$ mentioned before is justified here, since $K > K_c$, thus it allows for a synchronized state to exist.

As a measure of the vulnerability of each transmission line, we take advantage of the equation \eref{eq:phase_diff}, given that the edge $(i,j)$ is considered weak when $|\Delta_{\theta}(i,j)|$ is close to 1, while it is considered more stable when it is close to 0, regarding the synchronization limit imposed by \eref{eq:dorfler_sync_condition}. This approach is equivalent to that presented in \cite{Witthaut2012, Rohden_2017}, where the weakness of an edge is measured as the power transferred between the linked nodes (line load), except that here, the steady state of the system is approximated immediately from equation \eref{eq:phase_diff}.

\subsection{\label{sec:node_vulnerability} Nodal vulnerability}

Consider a multistable dynamical system that evolves in the state space $X$ and let $X^{*}\subset X$ be the set of desirable attracting states. The \emph{basin of attraction} of $X^{*}$, noted by $\beta$, is defined as the set of all initial conditions $x_{(0)}$ that asymptotically converge to $X^{*}$ \cite{Menck2013}. For the purpose of this work, $X^*$ will be the frequency synchronization manifold of the system \eref{eq:kuramoto}. Similarly, the likelihood of a randomly perturbed trajectory to return back to $\beta$, known as the \emph{basin stability} $S_{B}$, can be defined as \cite{Menck2013, Menck2014}:
\begin{equation}
S_{B (\beta)}=\int\Gamma_{\beta (x)} \rho_{(x)} dx\label{eq:bs}
\end{equation}
where $\Gamma_{\beta (x)}$ is a function that indicates whether a state $x$ belongs to the basin of attraction of $X^{*}$ or not, that is

\begin{equation}
\Gamma_{\beta (x)} = 
 \cases{
1& $\text{\ensuremath{\forall}}x\in\beta$\\
0& $\text{\ensuremath{\forall}}x\notin\beta$
}
\label{eq:gam}
\end{equation}

The function $\rho_{(x)}$ is the density of states to which the system can be pushed by some non-local perturbation, such that $\intop_{X}\rho_{(x)}dx=1$. Note that under this definition, the basin stability is a number between 0 and 1, being $S_{B}=0$ when the synchronous state is unstable and $S_{B}=1$ when it is globally stable.

Monte Carlo simulations can be performed in order to estimate $S_B$ by randomly sampling a sufficiently high number of disturbed states $I_{C}$ (following a certain distribution $\rho_{(x)}$) from a representative subspace $\Pi \subset X$, and using them as the initial conditions for time-evolution simulations. For this work, $\rho_{(x)}$ is chosen as an uniform distribution in the restricted subspace $\Pi$, that is:
\begin{equation}
\label{eq:rhodense}
\rho_{(x)}=
\cases{
\frac{1}{\left|\Pi\right| } & $\text{\ensuremath{\forall}}x\in\Pi$ \\
0 & $\text{\ensuremath{\forall}} x \notin \Pi$
}
\end{equation}

\begin{figure}
\begin{center}
\includegraphics[width=0.8\linewidth, trim=0 0.2 0 0, clip]{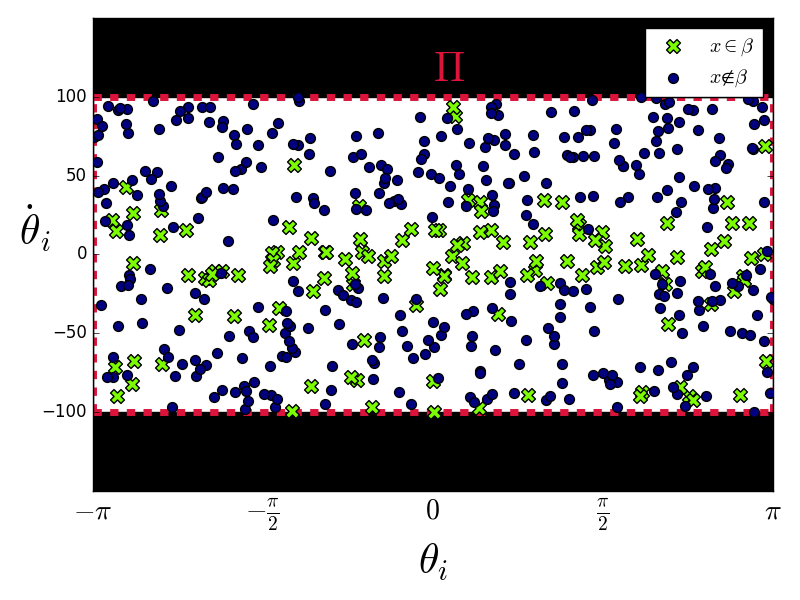}
\end{center}
\caption{Phase space of a node $i$ of the Colombian grid when random disturbances are applied to $\theta_i$ or $\dot \theta_i$. Green crosses are initial conditions that return to the frequency synchronized state, while the blue circles are out of the basin of attraction. The red dashed line denotes the subspace $\Pi$ from which disturbances are chosen. This diagram was created by sampling $I_C = 500$ points.}
\label{fig:MC_BS}
\end{figure}

Finally, the amount $F_{C}$ of simulated trajectories that are found to approach asymptotically to the attractor is assumed to be proportional to the volume of the basin of attraction (restricted to the subspace $\Pi$), thus $S_{B}$ is approximated by $S_{B}\approx\nicefrac{F_{C}}{I_{C}}$. Now, making use of the concept of basin stability we can define an indicator of the robustness of a node against large perturbations applied to it, by means of the single-node basin stability (SNBS) \cite{Menck2014, Mitra2017b}, given by:
\begin{equation}
\begin{array}{cc}
S_{B}^{(i)}=\frac{F_{C}}{I_{C}}, & i\in\vartheta\end{array}
\label{eq:stoch_bs}
\end{equation}
where the $I_{C}$ initial conditions are drawn from perturbations applied to node $i$. As illustration, figure \ref{fig:MC_BS} shows the phase plane of some node $i$ subject to large disturbances; its SNBS is, loosely speaking, the number of crosses divided by the number of total datapoints. Throughout this work, the disturbances for each node are drawn from the subspace $\Pi = [-\pi, \pi] \times [-100, 100]$.


\subsection{Structural vulnerability}

Equations \eref{eq:phase_diff} and \eref{eq:stoch_bs} define already a dynamical vulnerability for edges and nodes, respectively. Furthermore a vulnerability based merely on the connectivity features of the complex network could also be considered, since it is intuitive that by removing a highly connected node or a certain critically located edge could lead to a rapid diminishing on power grid stability and performance. In that regard, let us define structural vulnerability in terms of the following centrality measures:
\begin{itemize}
    \item \textit{Degree centrality} $d_k^{(i)}$: Amount of nodes to which node $i$ is connected, normalized by the maximum possible degree in the network. It can be computed~as:
\begin{equation}
d_k^{(i)} = \frac{1}{N-1} \sum_j^N{ a_{ij} }
\label{eq:degree}
\end{equation}
    \item \textit{Clustering coefficient} $c_k^{(i)}$: Number of triangles ($T_i$) in which node $i$ is involved, normalized by the maximum possible amount of such triangles \cite{Saramaki2007}, that is:
\begin{equation}
c_k^{(i)} = \frac{ T_i }{ d_k^{(i)}(d_k^{(i)} - 1) }
\label{eq:cluster}
\end{equation}
    \item \textit{Node betweenness centrality} $b_k^{(i)}$: Sum of the shortest paths between every pair of nodes $(s,t)$ in the network that pass through node $i$:
    
\begin{equation}
b_k^{(i)} = \sum_{s \neq t \neq i \in \vartheta} \frac{\sigma_{s,t}(i)}{\sigma_{s, t}}
\label{eq:betwnode}
\end{equation}

where $\sigma_{s, t}$ is the amount of shortest paths between nodes $s$ and $t$ and $\sigma_{s,t}(i)$ is the number of those paths that pass through node $i$ \cite{ulrikb}. 

    \item \textit{Edge betweenness centrality} $e_k^{(l)}$: In the same fashion as $b_k$ measures centrality for a node, $e_k$ does it for an edge; it is simply defined as the number of shortest paths in the network that include edge $l$:

\begin{equation}
e_k^{(l)} = \sum_{s \neq t \in \vartheta} \frac{\sigma_{s,t}(l)}{\sigma_{s, t}}, ~~~ l \in \varepsilon
\label{eq:betwedge}
\end{equation}

\end{itemize}

So in this work, a node will be considered structurally weak if $d_k$, $c_k$ or $b_k$ is high. Similarly, an edge will be considered structurally weak if its $e_k$ is high.

\section{\label{sec:resilresul}Resilience measures}

In order to assess the resilience of a power grid, we propose a percolation-based algorithm. To do so, we will follow the capability of the network to maintain its functioning upon different node removal (site percolation) and edge removal (bond percolation) rules. In particular, the algorithm that will be described below removes a node or an edge from the graph on each iteration by applying a random attack, where the element to be removed is chosen uniformly at random; or a  focal attack, where the most vulnerable node or edge in the graph is removed first. Here, the most vulnerable node is the one with the lowest basin stability ($S_B^{(i)}$) or the highest degree centrality ($d_k$), clustering coefficient ($c_k$) or node betweenness centrality ($b_k$); while the most vulnerable edge is the one for which phase difference ($\Delta_{\theta}(i,j)$) or edge betweenness centrality ($e_k$) is the highest.

\begin{figure*}
\includegraphics[width=0.98\linewidth]{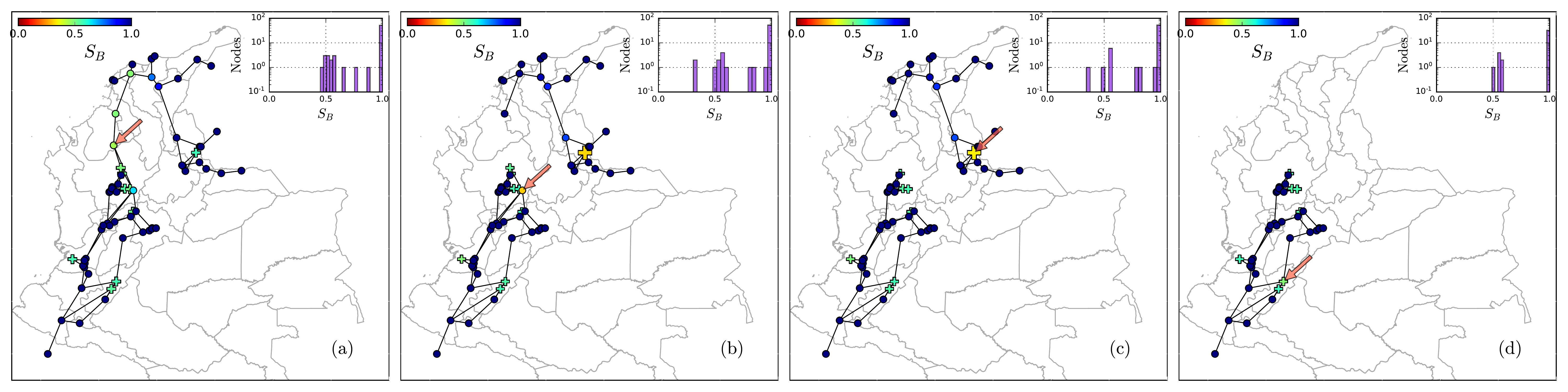}\\
\includegraphics[width=0.98\linewidth]{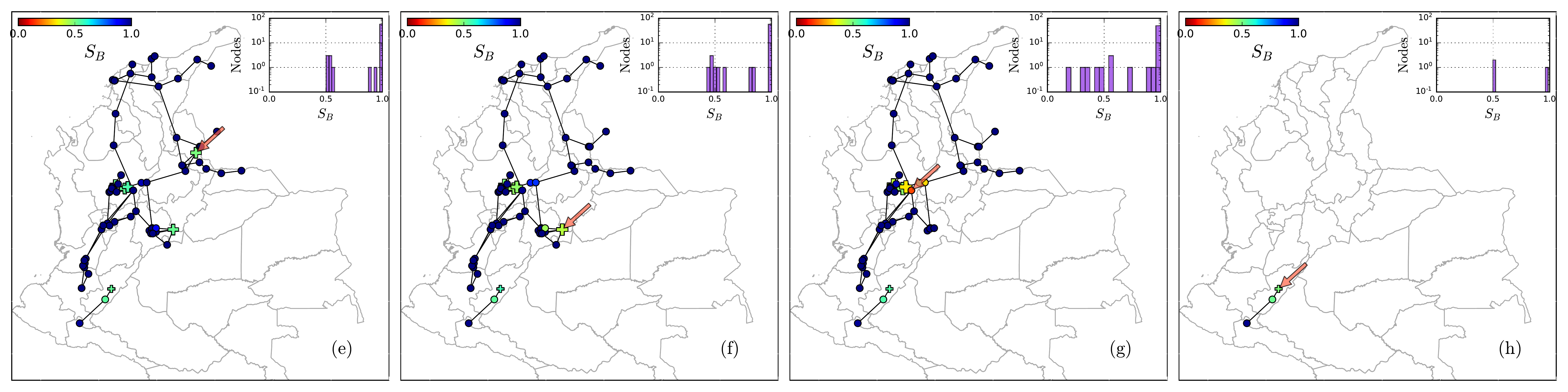}
\caption{Node removal algorithm under the $S_B$-focused attacking scheme. Circle nodes indicate consumers while crosses represent generators. Node color is mapped with the $S_B^{(i)}$ (computed with $I_C = 50$) and the size of the node is proportional to the power $P_i$. The red arrow indicates the node that is going to be suppressed. Also, the inset shows the histogram of the basin stability for the whole network. One realization is presented in panels (a)-(d) and a different one in panels (e)-(h) in order to visualize two different mechanisms of cluster vanishing.
\label{fig:nodal_algorithm}}
\end{figure*}

The percolation-based algorithm then proceeds as such:

\begin{enumerate}
\item Remove a node (edge) by applying a random or a focal attack. Note that in a focal attack, multiple nodes (edges) could share the same vulnerability level, in that case, the attacked node is chosen randomly among them.

\item Compute the existing disconnected clusters in the new graph.

\item For each cluster, check if it contains at least one generator node and one consumer node, if it does not, remove the whole cluster from the graph.

\item For each remaining cluster the power has to be compensated, so as a way to simulate a real-world power control (where the supplied power level is increased or reduced according to the demand of the consumers), the power of all generators in the cluster is modified uniformly such that equation \eref{eq:powerbalance} remains true.  

\item For each remaining cluster, check whether or not it is synchronizable, that is, its topology satisfies the synchronization condition that $K > K_c$, where $K_c$ is computed by the approximation \eref{eq:Dorfler_critical}. If it does not, the whole cluster is removed from the graph.

\item Measure the needed observables and return to step 1 until the whole power grid has gone to blackout, that is, no operational cluster exists in the network.
\end{enumerate}

An important remark has to be done about the algorithm, regarding node removal under the focal attack methodology based on SNBS. Computing this vulnerability index is computationally expensive since it requires a large amount of time-domain simulations in order to estimate reasonably well the equation \eref{eq:stoch_bs}. Thus, in the following, for the $S_B$-focused percolation, only 10 simulations are considered for averaging purposes, while for each of the other cases, 500 simulations are performed. It is also worth noting that the fine structure of the basin of attraction is not captured by this approach, and it could impose numerical problems on dynamical systems that posses either fractal basin boundaries or riddled or intermingled basins \cite{Schultz2017}.

\begin{figure*}
\includegraphics[width=0.98\linewidth]{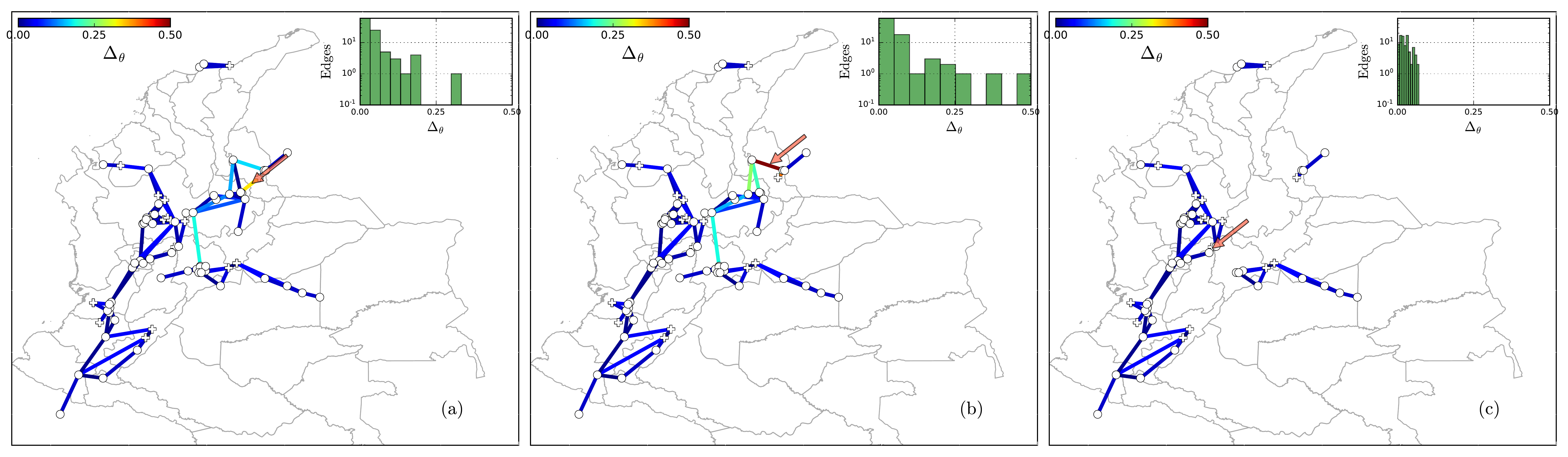}
\caption{Edge removal algorithm under the focal attacking scheme. Edge color is mapped with the $\Delta_{\theta}$ of the nodes it connects. The red arrow indicates the edge that is going to be supressed. Also, the inset shows the histogram of the phase differences for the whole network.
\label{fig:link_algorithm}}
\end{figure*}

For this elimination procedure, two main phase transitions will be analysed:
\begin{itemize}
\item \textbf{Transition $\boldsymbol{T_1}$:} This transition occurs at the iteration when the graph is divided into multiple functional subgraphs for the first time (turning from a connected graph to an unconnected graph).

\item \textbf{Transition $\boldsymbol{T_2}$:} The final iteration of the algorithm, when the power grid goes to complete blackout.\\ 
\end{itemize}
Figure \ref{fig:nodal_algorithm} illustrates the algorithm for nodal resilience measurement under focal attacks. The transition from the frame (a) to (b) shows the formation of 2 independent clusters after removing one node. Note that in the frame (c) there is only one generator in the northern cluster, and interestingly, it has a very poor basin stability level, so when that node is removed from (c) to (d), the whole northern cluster goes to blackout. A different realization is presented in panels (e)-(h); note that from (g) to (h) it suffices to remove one node to bring the bigger cluster to blackout even though there would still remain connected subgraphs with both generators and consumers. This occurs because the remaining cluster would have a topology that does not sattisfy the synchronization condition \eref{eq:dorfler_sync_condition} and thus it vanishes. Similarly, figure \ref{fig:link_algorithm} shows the algorithm in action under $\Delta_\theta$-focused attacks. In the transition from panel (b) to (c), it can be appreciated that, after suppressing a very vulnerable transmission line, a significant cluster of isolated consumers in the east side of the network vanishes.

\begin{figure*}
\centering
\subfloat[Giant component of the network.]{\includegraphics[width=0.5\linewidth]{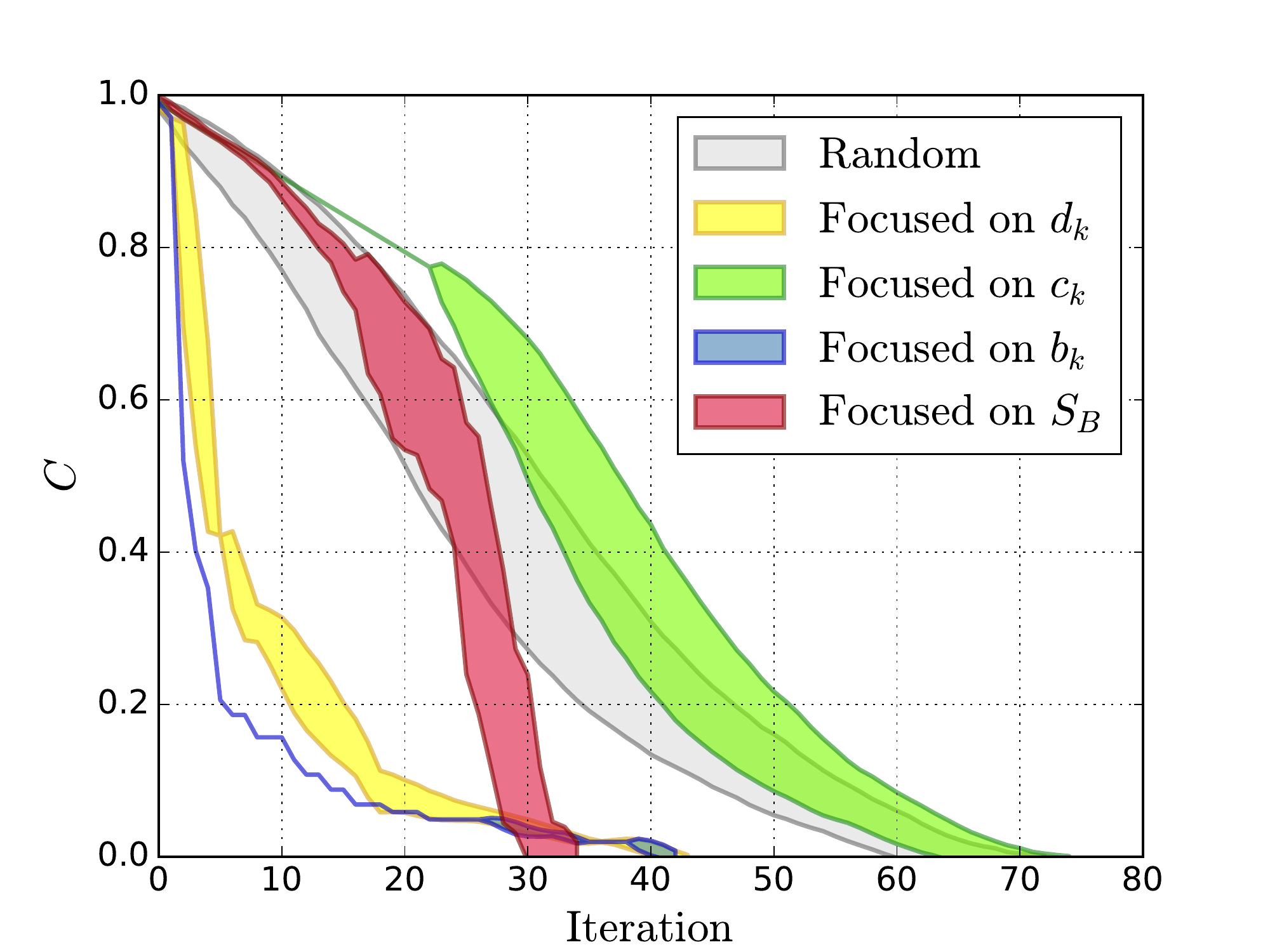} }
\subfloat[Removed component of the network.]{\includegraphics[width=0.5\linewidth]{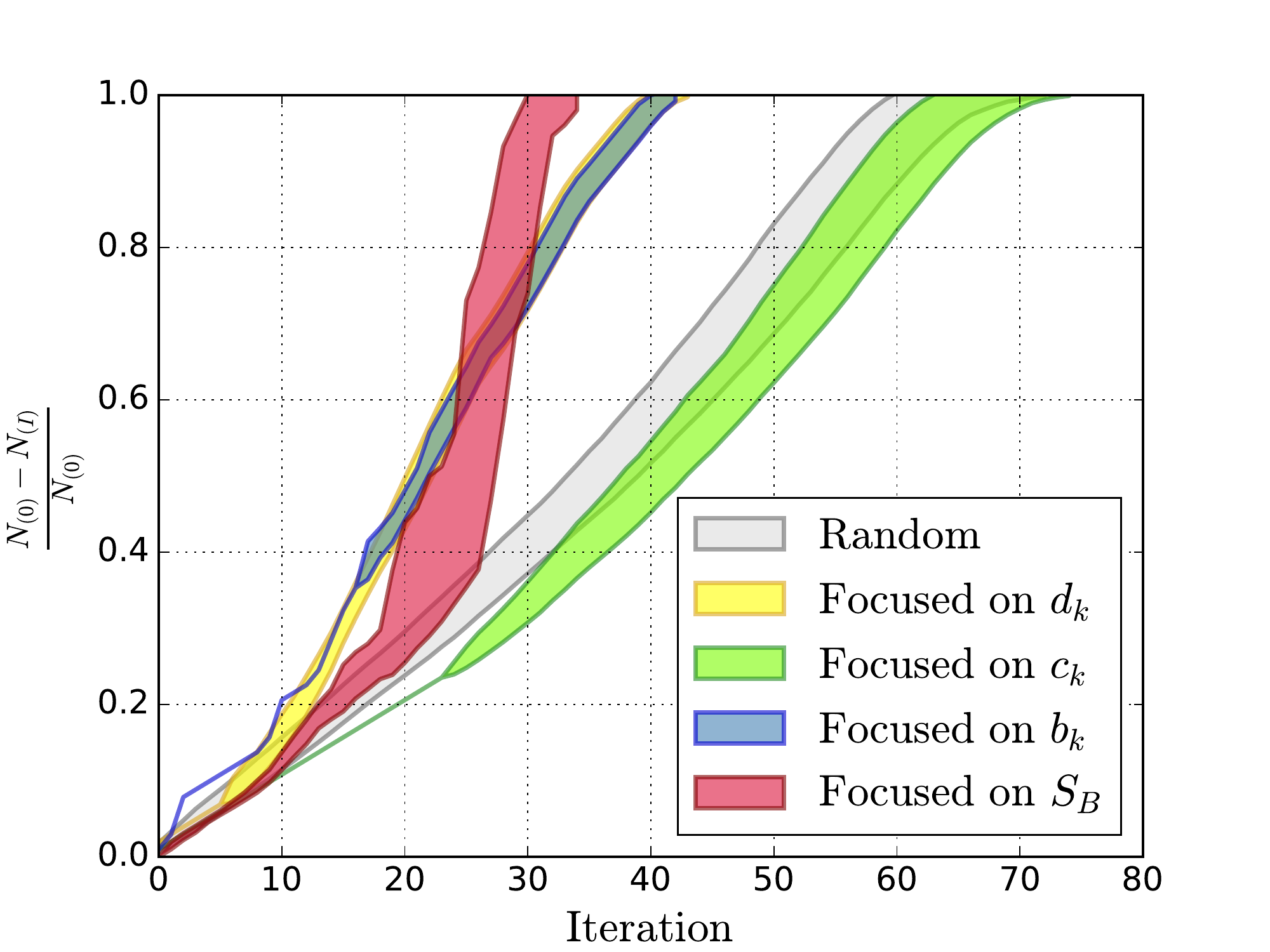}}\\
\subfloat[Second largest component of the network.]{\includegraphics[width=0.5\linewidth]{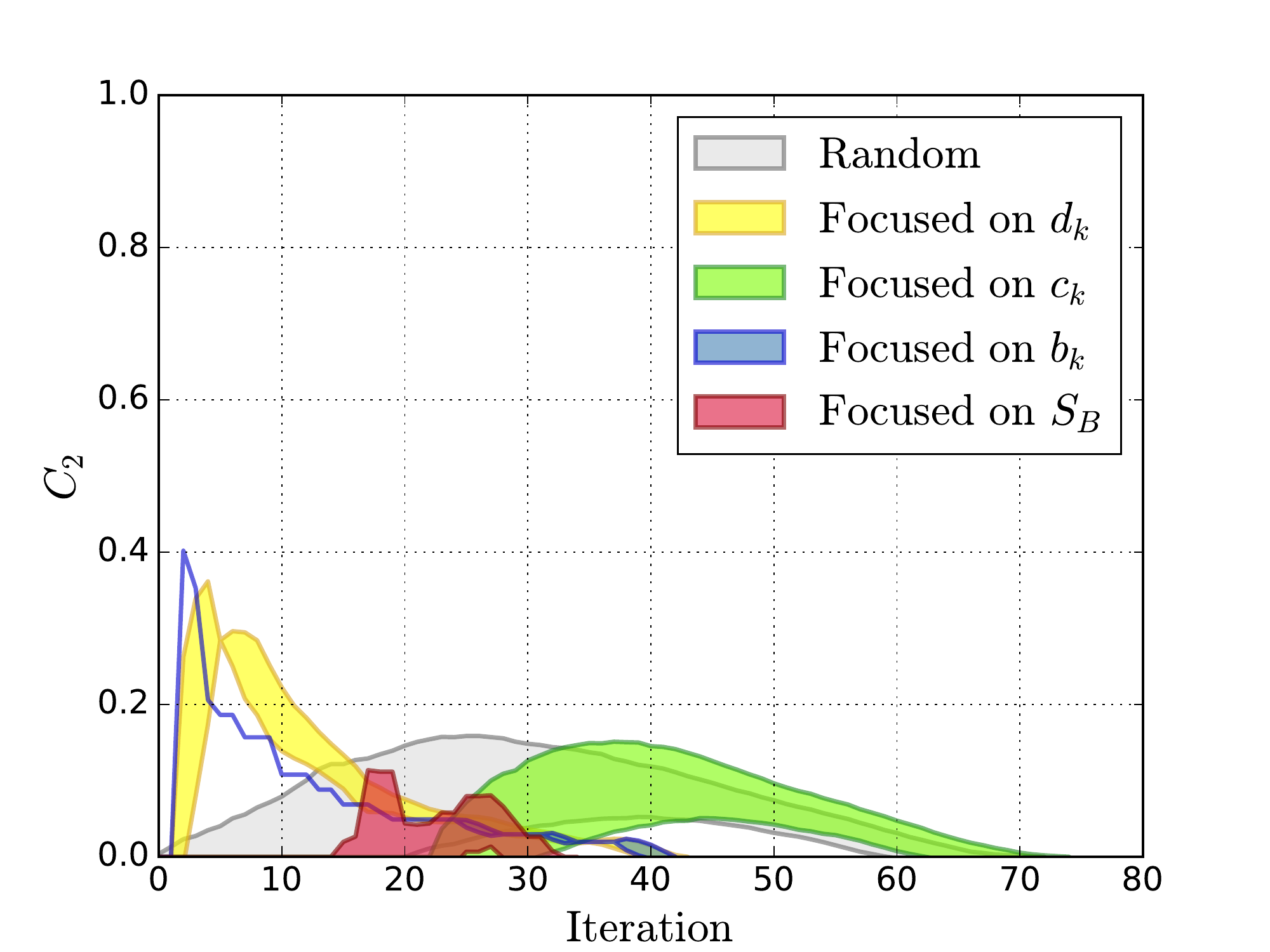}}
\subfloat[Amount of operational clusters in the network.]{\includegraphics[width=0.5\linewidth]{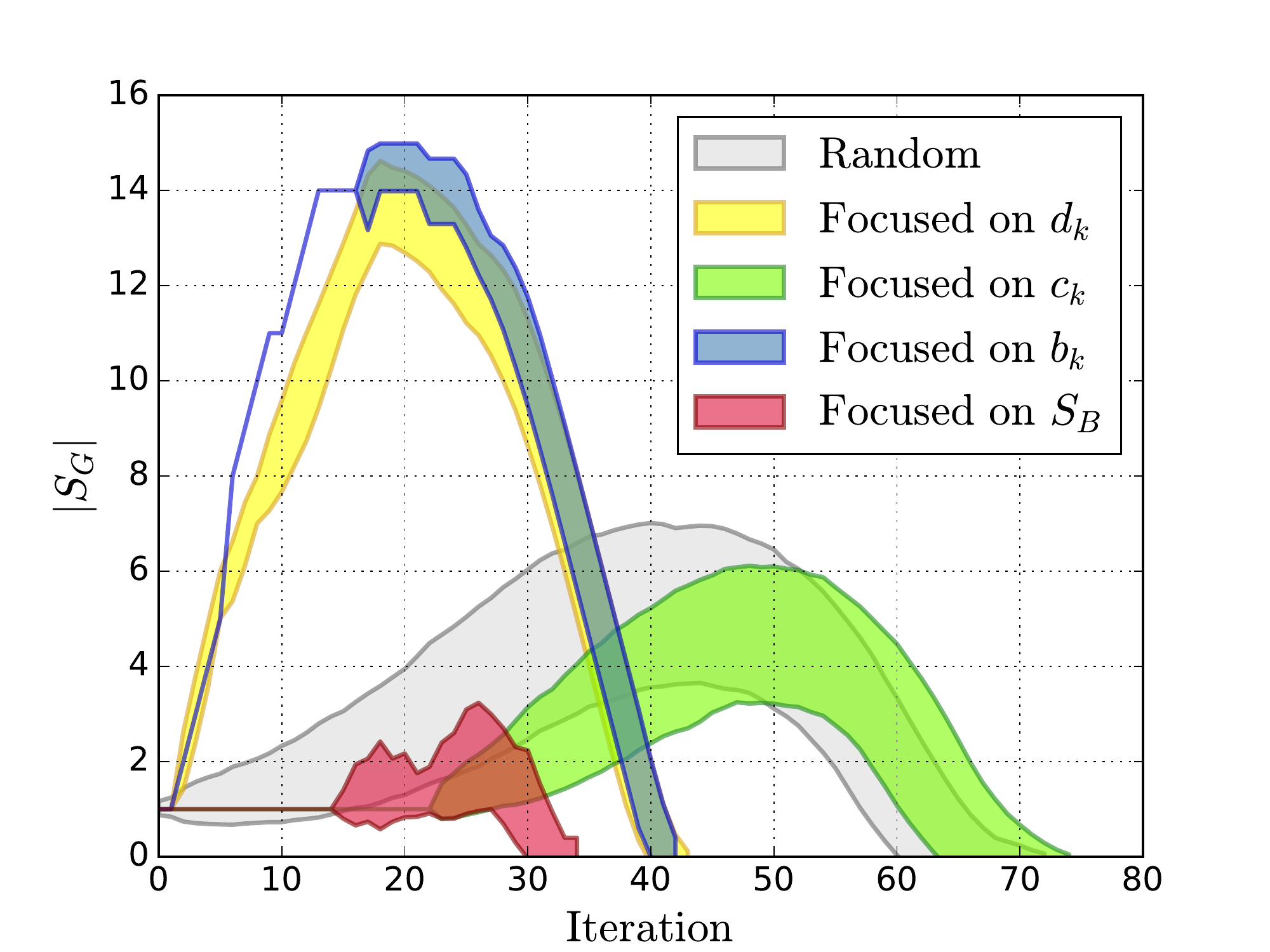}}
\caption{Evolution in the structure of the \textbf{Colombian power grid} subject to the nodal elimination algorithm. The shading accounts for the standard deviation calculated over 10 realizations with $I_C = 50$ for the $S_B$-focused methodology and 500 realizations for the other attacking schemes. 
\label{fig:node_percolation}}
\end{figure*}

\subsection{Node removal resilience for the Colombian power grid}

Figure \ref{fig:node_percolation}(a) shows the fraction of nodes in the largest cluster of the network (also called the giant component of the graph and denoted by $C$) subject to node removal with focal and random attacking policies. Similarly, figure \ref{fig:node_percolation}(b) presents the fraction of removed nodes after each iteration with respect to the original size (named removed component and denoted by $\nicefrac{(N_{(0)} - N_{(I)})}{N_{(0)}}$, being $N_{(0)}$ and $N_{(I)}$ the original size of $G$ and its size after $I$ removal iterations, respectively), which allows to detect the phase transition $T_2$, corresponding to the iteration at which $\nicefrac{(N_{(0)} - N_{(I)})}{N_{(0)}} = 1$. It is thus observed that the grid goes to complete blackout around 30th removal iteration for the case of the focal attacking on $S_B$, and around the 62nd iteration for the random attacking. For $d_k$ and $b_k$ the transition occurs at the 40th iteration and for the random and $c_k$ cases, we have the less effective removal technique since this transition occurs around the 60th attack.

The fact that attacking the most vulnerable nodes based on $S_B$ produces a faster transition to the total blackout state of the network than just removing nodes at random implies that there is indeed a relationship between the stability of a node against dynamical disturbances (related to the SNBS) and the structural connectivity and resilience to cascading failures of the graph.
\begin{figure*}
\centering
\subfloat[Giant component of the network.]{\includegraphics[width=0.5\linewidth]{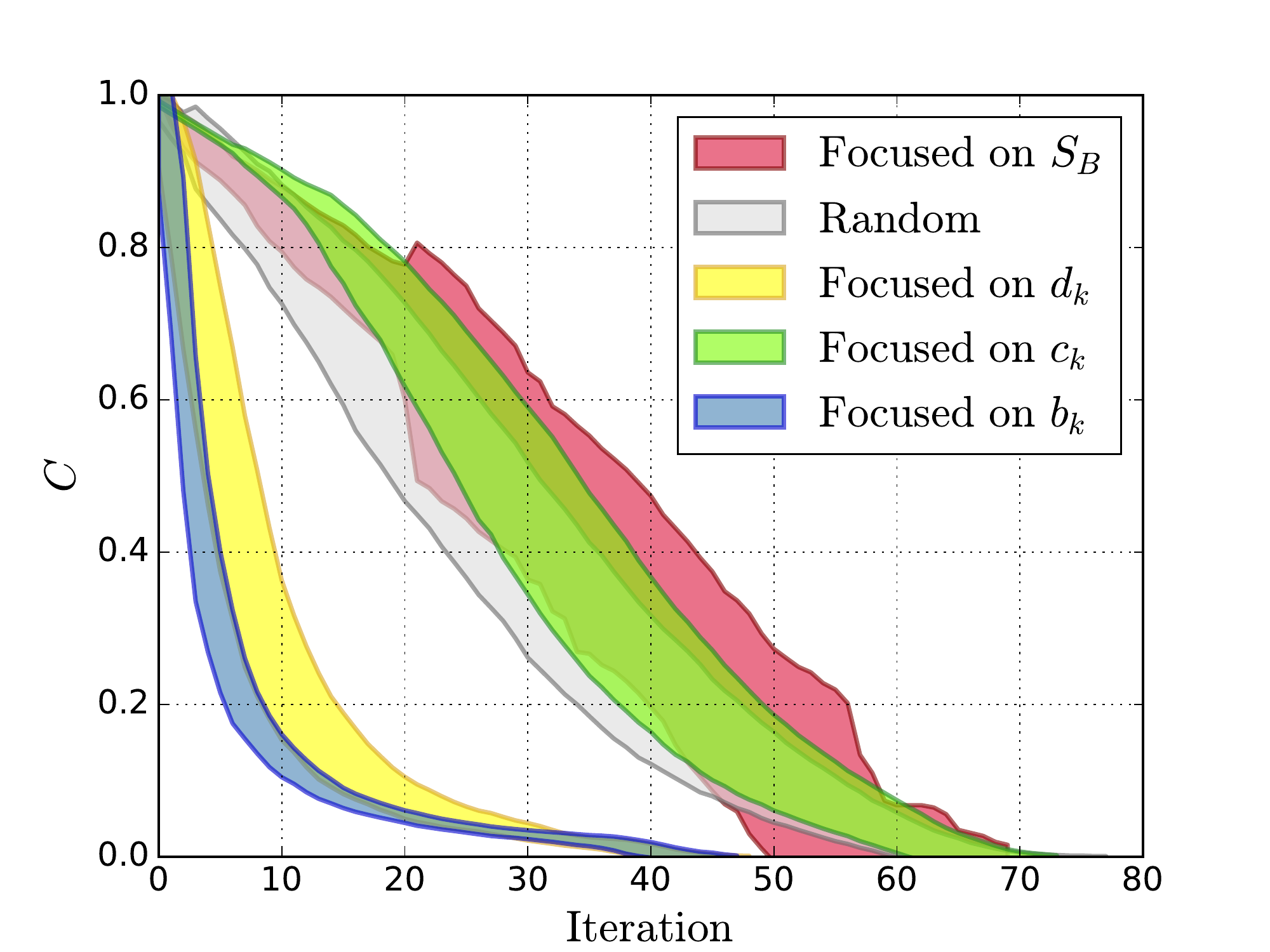}}
\subfloat[Removed component of the network.]{\includegraphics[width=0.5\linewidth]{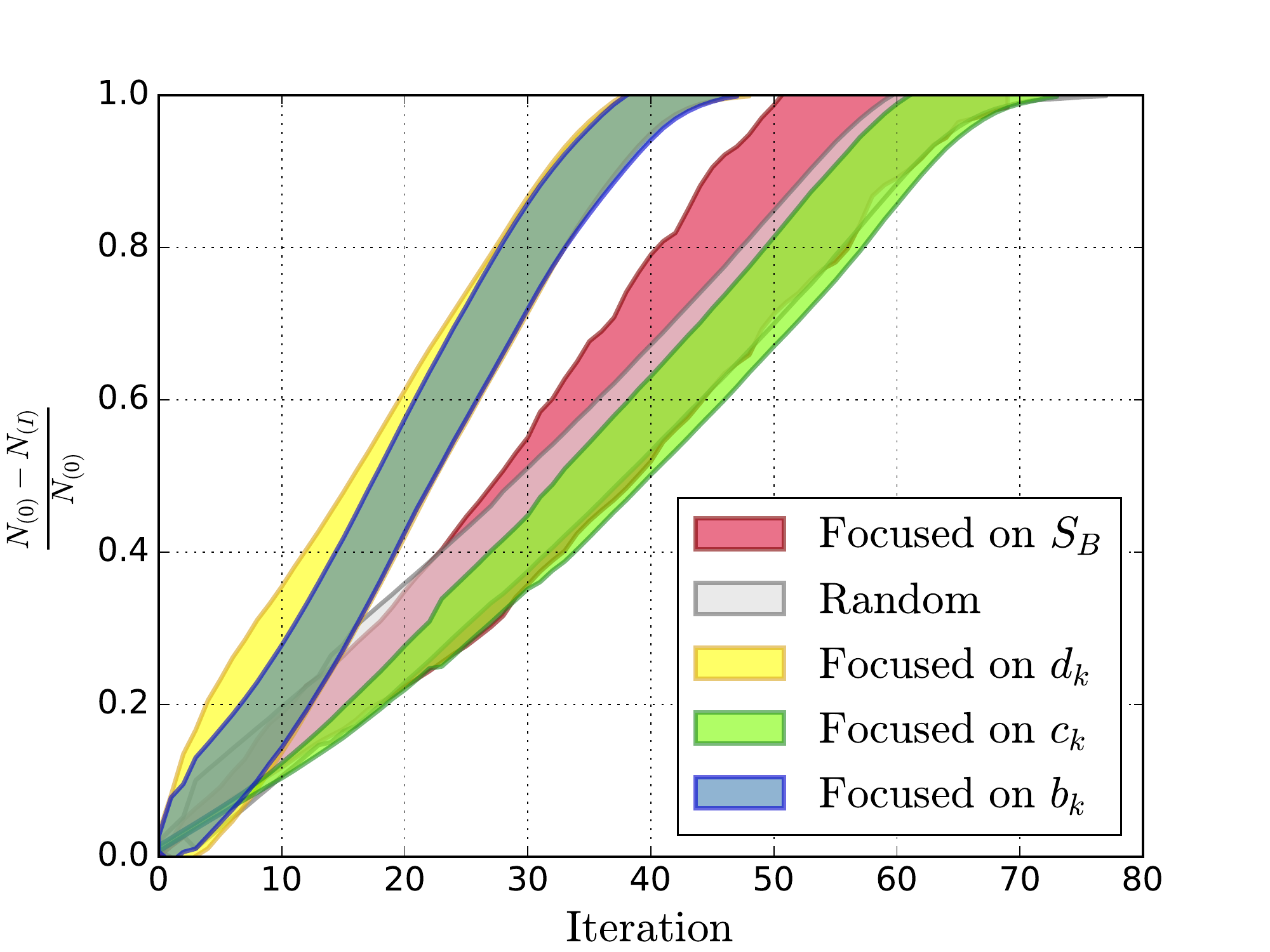}}\\
\subfloat[Second largest component of the network.]{\includegraphics[width=0.5\linewidth]{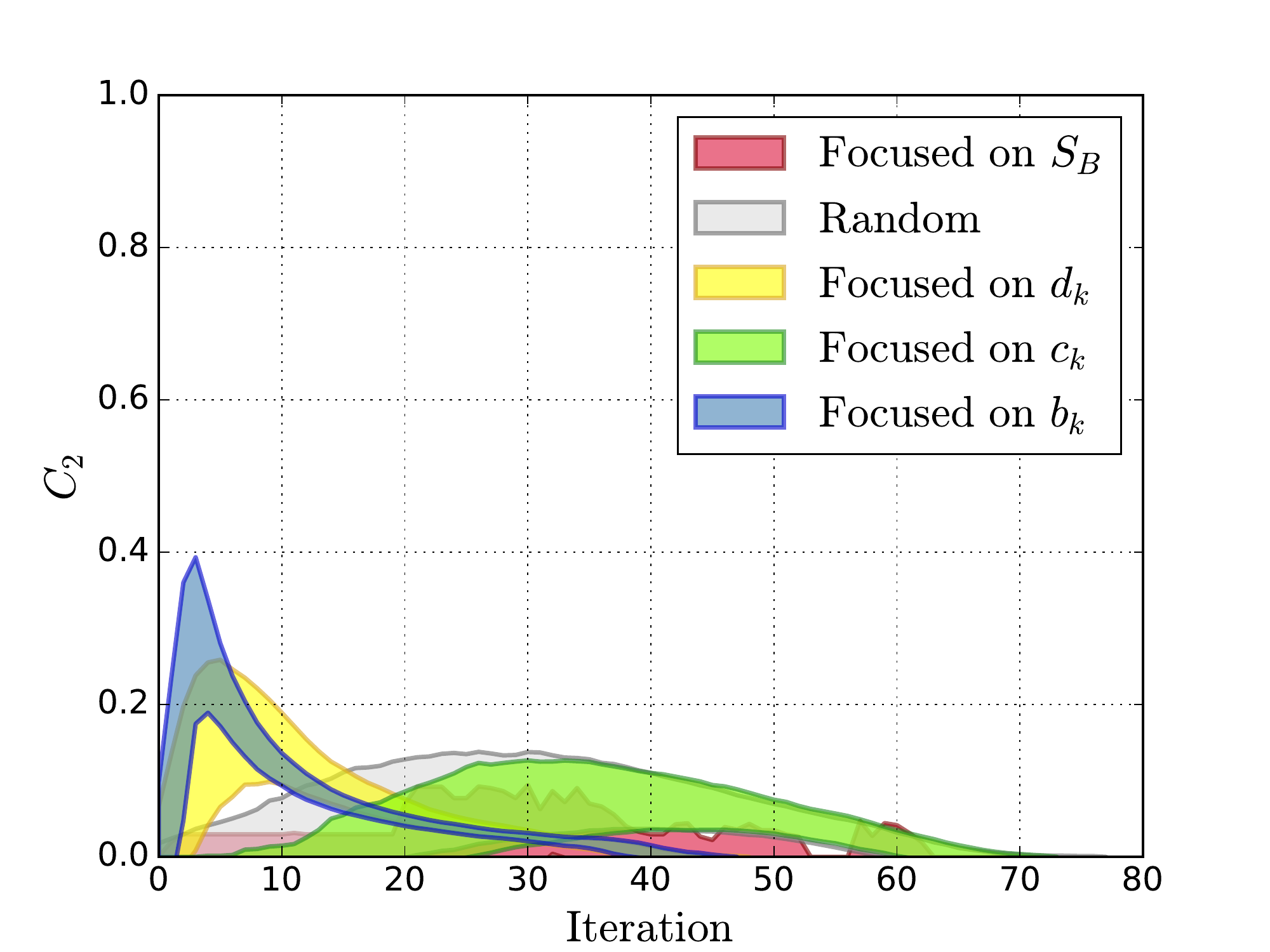}}
\subfloat[Amount of operational clusters in the network.]{\includegraphics[width=0.5\linewidth]{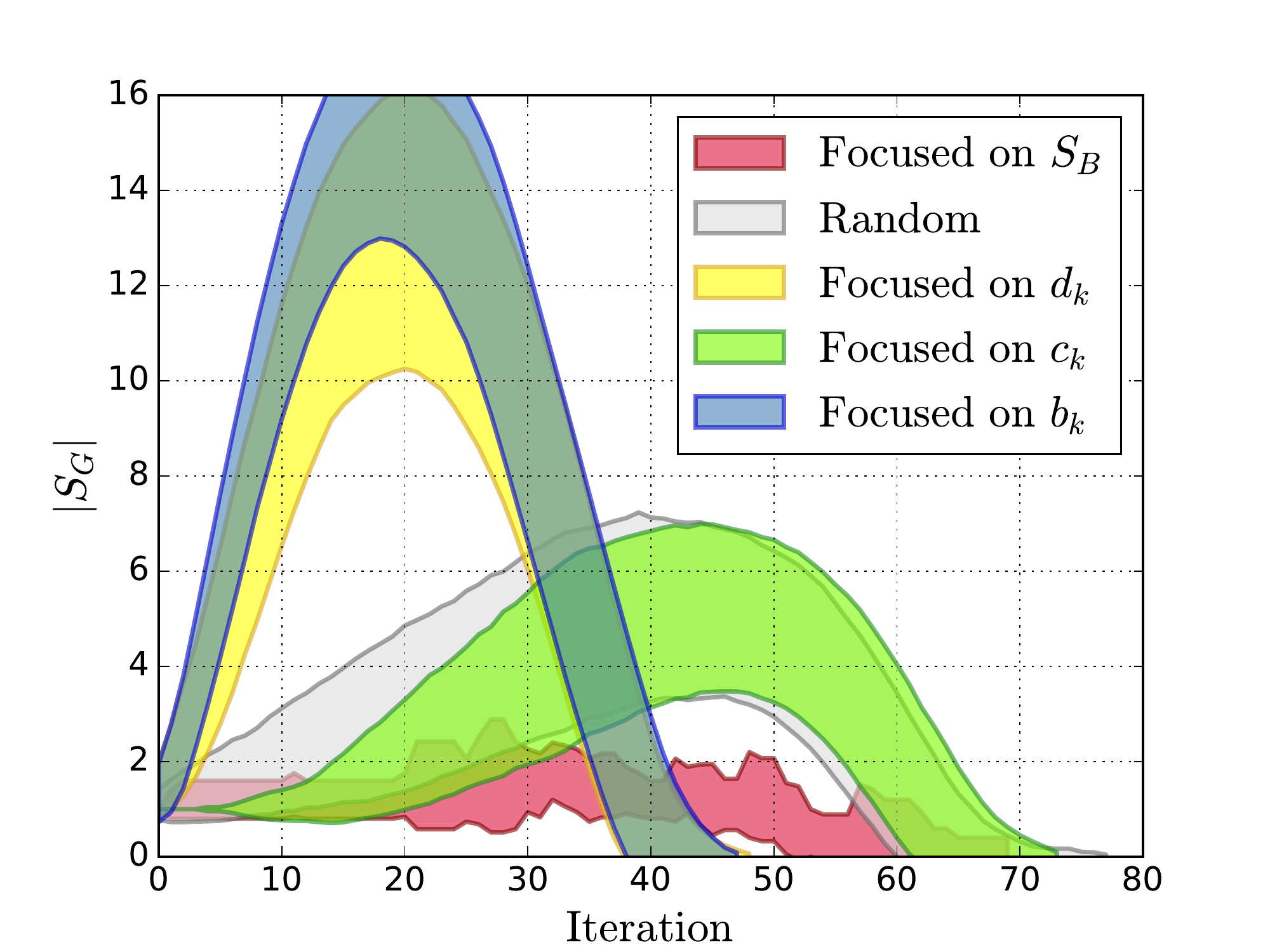}}
\caption{Evolution in the structure of the \textbf{randomly grown synthetic power grids} subject to the nodal elimination algorithm. The shading accounts for the standard deviation calculated over 10 realizations with $I_C = 50$ for the $S_B$-focused methodology and 500 realizations for the other attacking schemes.
\label{fig:node_percolation_schultz}}
\end{figure*}

Furthermore, by focusing the attacks on nodes with the highest $b_k$ or $d_k$, the dimension of the giant component goes down drastically during the first elimination iterations and then it goes down at a slower rate, reaching total blackout in an intermediate level of attacks between the $S_B$ and random methodologies. $c_k$ on the other hand, produces a slower reduction of the graph size and a total blackout with the most iterations in average. This is related to the tree-like structure of the power grid that can be seen on figure \ref{fig:ColombianPowerGrid}. In this kind of topologies, the nodes with the higher degree and betweenness are located at the bulk regions of the graph, therefore, their removal likely divides the giant component instantly into multiple smaller clusters, as also confirmed by the figure \ref{fig:node_percolation}(d), where it is shown that these methodologies produce by far, the highest amount of operational clusters (quantity denoted by $|S_G|$) during the whole procedure. Correspondingly, a node with a high clustering coefficient, by definition, implies that the neighbours that it is connected to, are also connected between them, hence, removing it is not going to divide the graph and generate new clusters. Eventually, after multiple node eliminations, redundant links between said neighbours will be suppressed and $c_k$ will become uniform over the whole system. That is the reason why this attacking strategy behaves roughly in the same way as the random attacking one after a few iterations.

Figure  \ref{fig:node_percolation}(c) plots the size of the second largest component of the graph $C_2$, which reveals the location of the phase transition $T_1$ as the first iteration when $C_2$ jumps from 0 to a non-zero value. This figure shows, as expected, that the $d_k$ and $b_k$ attacking policies generate in a couple of iterations a new cluster composed by a significant amount of nodes (peaking roughly at 40\% of the nodes), while random and $c_k$ focal methods generate smaller subgraphs and the transition takes in average, a higher attacking effort. Interestingly, the transition to an unconnected graph composed of multiple functioning clusters for the $S_B$ case, takes several iterations (at least 15 iterations) and it occurs closer to the total blackout transition (about 30 iterations) than any other methodology.

This behavior is further complemented with figure  \ref{fig:node_percolation}(d), here, the $S_B$ focal strategy generates very few clusters as compared to the others; this is likely related again to the tree-like topology of the Colombian power grid and to the observed phenomenon in \cite{Menck2014}, where nodes located on dead-tree arrangements (more specifically, inner tree nodes \cite{Nitzbon2017}), were found to exhibit, in general, a lower SNBS than the rest of the nodes. As these are some of the nodes that are more likely to be removed first under $S_B$ focal attacks, and as by definition they do not have a significant amount of connections, the graph does not segregate on multiple clusters that easily.

Table \ref{tab:t1t2_col} summarizes the average value of the transitions for the node removal algorithm applied to the Colombian power grid.

\begin{table}[h]
\centering
\caption{\label{tab:t1t2_col}Average value of $T_1$ and $T_2$ for the nodal resilience testing algorithm in the Colombian grid. }
\begin{tabular}{|c|c|c|c|c|c|} 
\hline
\textbf{} & \textbf{Random Node} & $\boldsymbol{b_k}$ & $\boldsymbol{c_k}$ & $\boldsymbol{d_k}$ & $\boldsymbol{S_B^k}$  \\ 
\hline
$\boldsymbol{T_1}$ & 14.41 & 3.00 & 28.77 & 3.18 & 20.60 \\
\hline
$\boldsymbol{T_2}$ & 61.92 & 40.85 & 65.03 & 40.54 & 29.80 \\
\hline
\end{tabular}
\end{table}

\subsection{Node removal resilience for synthetic power grids}

To test the generality of the results shown for the Colombian power grid, the process is repeated for a set of synthetic power grids, which are generated as described in section \ref{sec:powerGrid}. Results on the nodal resilience for these synthetic power grids are presented in figure \ref{fig:node_percolation_schultz}. It is surprising that, besides the wider standard deviation in every trace (which was to be expected considering the randomness in the generation of the complex network), every other behavior related to the elimination based on $b_k$, $c_k$, $d_k$ and even the random attacking method is identical to that observed in the specific case of the Colombian power grid in figure \ref{fig:node_percolation}. This confirms the fact that the algorithm proposed in \cite{Schultz2014} is indeed capable of reproducing real-world examples of networked power systems, but also reveals that an homogeneous power distribution did not have, in average, a significant impact in the transition to blackout for these specific attacking schemes. This is not surprising, as $b_k$, $c_k$ and $d_k$ are structure-based vulnerabilities and completely independent of the dynamical vulnerabilities. The $S_B$-focused attacking, however, shows a very different evolution, similar to that of the random attacks, except in figure \ref{fig:node_percolation_schultz}(d), where $S_B$ attacks keep producing the lowest amount of clusters, which is related to the same explanation given before: the nodes located at dead-tree arrangements usually possess low SNBS levels, and their removal hardly generates new functional clusters. For reference, the values for the transitions $T_1$ and $T_2$ on the synthetic power grids are presented in table~\ref{tab:t1t2_shultz}.

\begin{table}[h]
\centering
\caption{\label{tab:t1t2_shultz}Average value of $T_1$ and $T_2$ for the resilience testing algorithm in the synthetic power grids. }
\begin{tabular}{|c|c|c|c|c|c|} 
\hline
\textbf{} & \textbf{Random Node} & $\boldsymbol{b_k}$ & $\boldsymbol{c_k}$ & $\boldsymbol{d_k}$ & $\boldsymbol{S_B^k}$  \\ 
\hline
$\boldsymbol{T_1}$ & 10.20 & 2.35 & 19.58 & 2.50 & 27.00 \\
\hline
$\boldsymbol{T_2}$ & 61.62 & 39.91 & 63.22 & 39.02 & 54.30 \\
\hline
\end{tabular}
\end{table}

It seems quite natural to think that, as the random networks have more generator nodes and they are randomly distributed over the graph, the iteration for which the transition to blackout occurs would, for any of the attacking methods, be higher than the Colombian case since after each removal step the probability to obtain non-operational clusters would be reduced, yet it only happens for the $S_B$-focused one, as can be also confirmed by comparing tables \ref{tab:t1t2_col} and \ref{tab:t1t2_shultz}. Next section focuses the discussion around the effects of these power heterogeneity in the resilience profiles observed.

\subsection{Influence of power heterogeneity}

\begin{figure}[h]
\begin{center}
\subfloat[Colombian power grid.]{\includegraphics[width=0.5\linewidth]{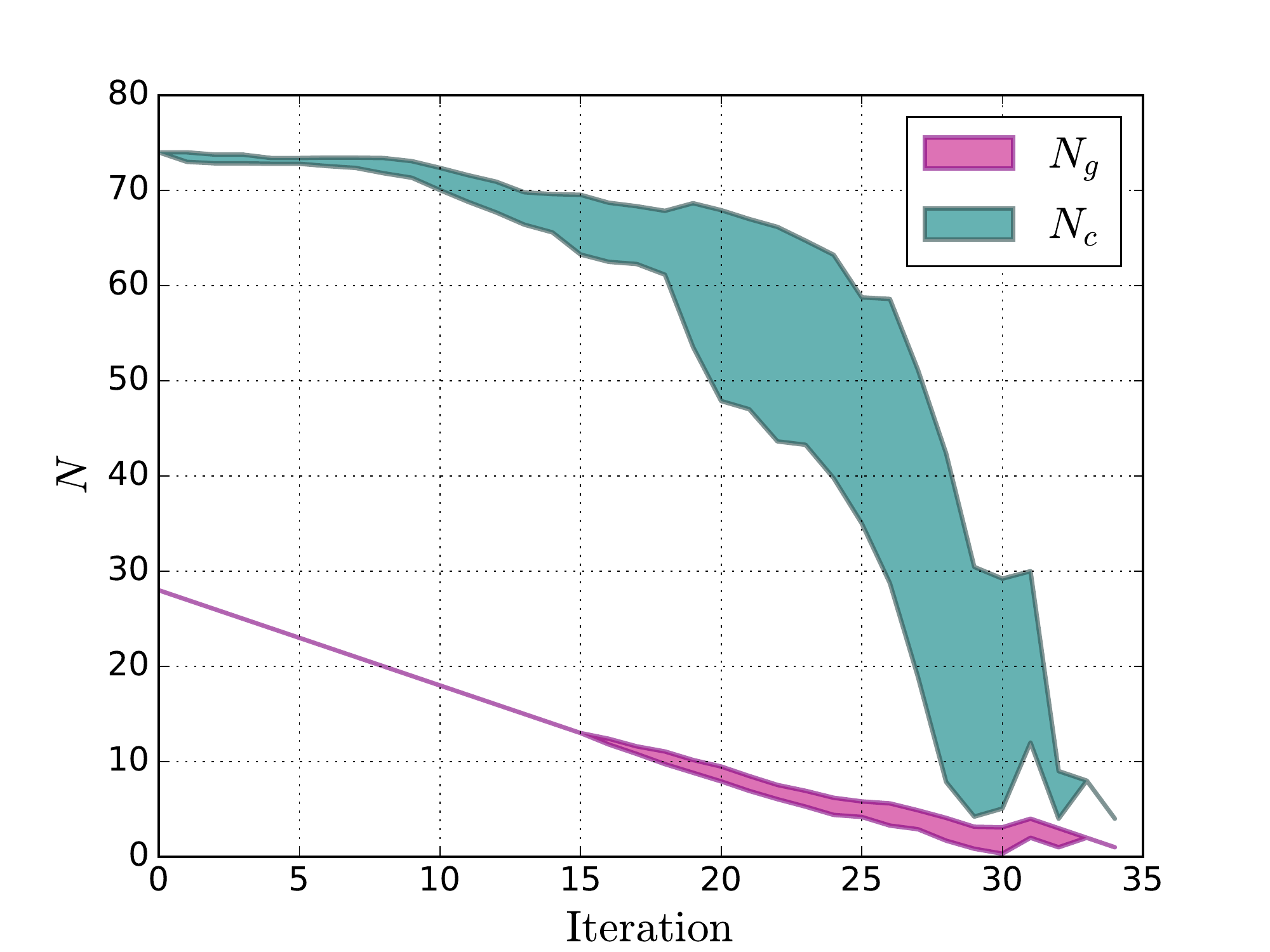}}
\subfloat[Synthetic power grids.]{\includegraphics[width=0.5\linewidth]{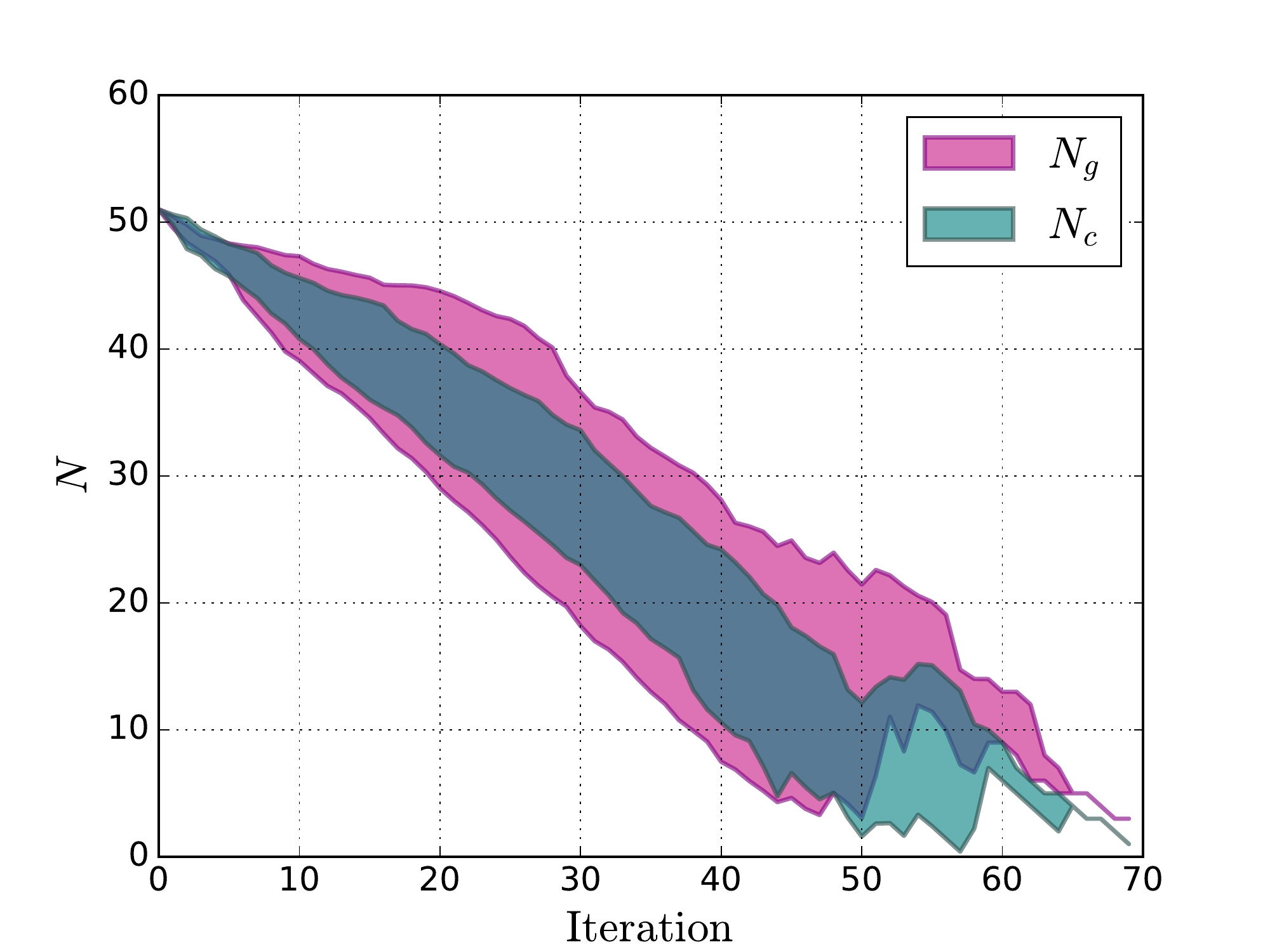}}
\end{center}
\caption{Number of consumers $N_c$ and generators $N_g$ existing in the graph on each iteration of the $S_B$-focused removal algorithm. The shading accounts for the standard deviation over the 10 random realizations with $I_C = 50$.
\label{fig:rem_nodes}}
\end{figure}

Figure \ref{fig:rem_nodes} shows the amount of generators $N_g$ and consumers $N_c$ after each iteration in both, the Colombian power grid and the synthetic power grids, and it allows to see how the node elimination is working: for the random synthetic power grids in average, both, consumers and generators are being eliminated at the same rate, this leads to believe that every new cluster created in the network preserves a comparable number of both kind of nodes and that homogeneity makes the total destruction of said cluster less likely. This also explains why, in average, $T_2$ for the $S_B$-method is close to that of the random case (table \ref{tab:t1t2_shultz}) and hints that the main reason why clusters are being removed from the graph is because the synchronization condition is not satisfied (step 5 in the resilience testing algorithm), rather than the absence of generators and consumers (step 3 of the algorithm). Nevertheless, for the Colombian power grid the decreasing slope for $N_g$ is faster than that of $N_c$ for the first iterations. Since the initial amount of generators is lower than the initial amount of consumers in this test case, after some of the nodes in the former group have been removed from the graph, the resulting clusters are more likely to be consumers isolated without a connected generator, subsequently, these clusters go to blackout. This implies that the main reason why clusters are being suppressed from the Colombian grid is the absence of generators in them (step 3 of the algorithm), and explains why $T_2$ in the $S_B$-focused method is the lowest value in table~\ref{tab:t1t2_col}.

\begin{figure*}
\centering
\subfloat[Colombian power grid.]{\includegraphics[width=0.5\linewidth]{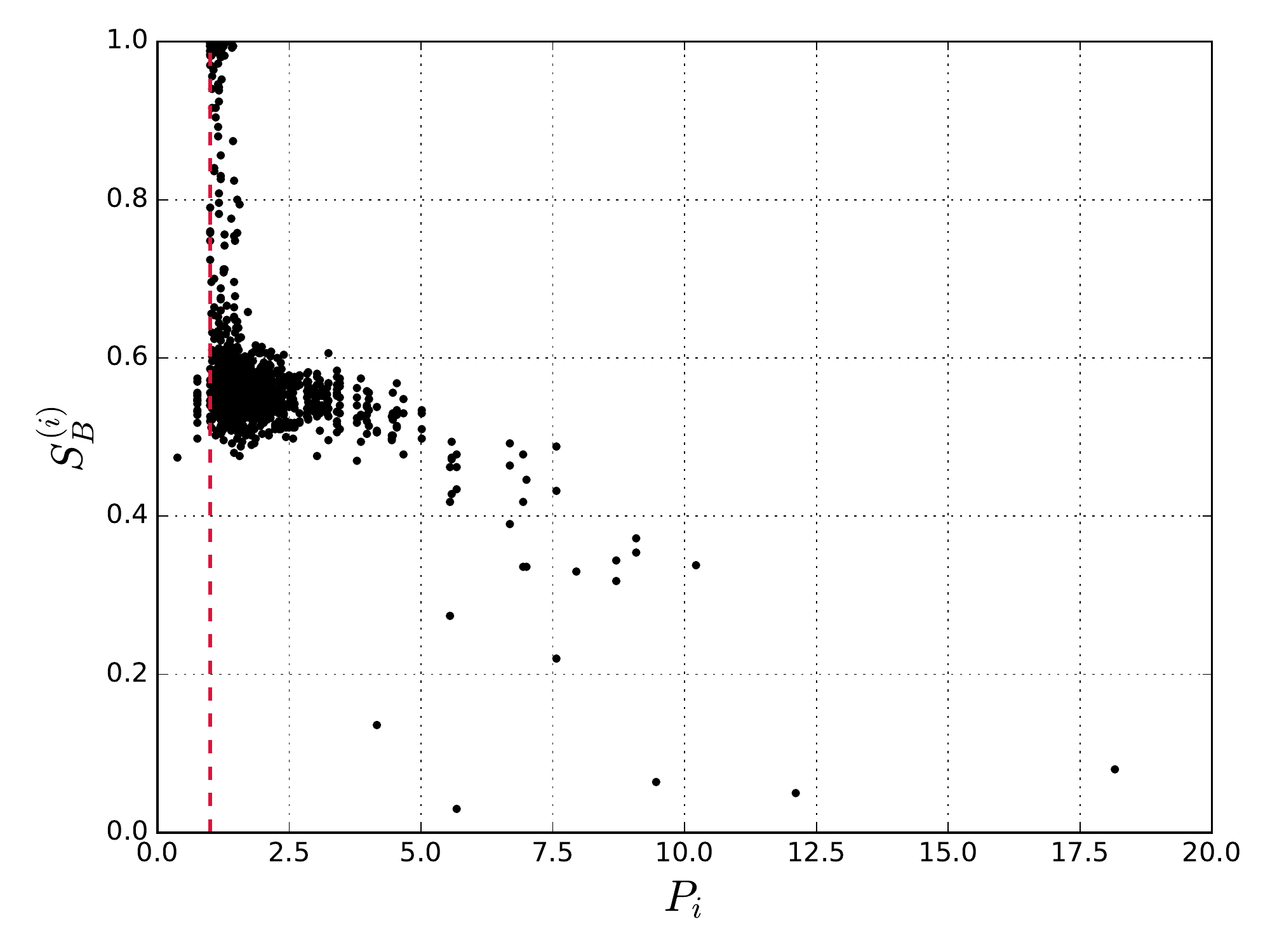}}
\subfloat[Synthetic power grids.]{\includegraphics[width=0.5\linewidth]{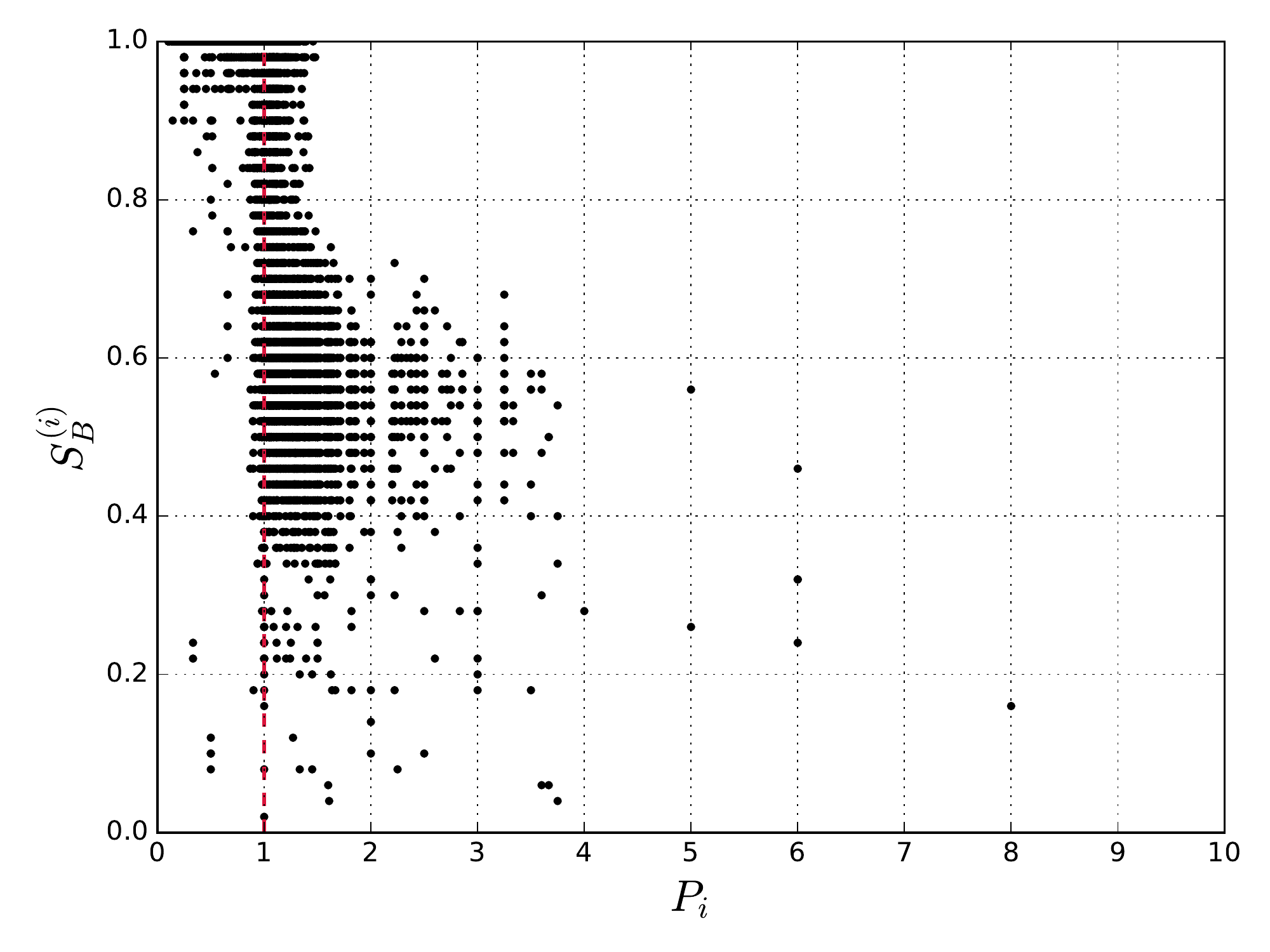}}
\caption{SNBS of each generator node as a function of the power it supplies to the network. Red dashed line marks the initial set power $P=1.0$.
\label{fig:sb_of_p}}
\end{figure*}

In addition, as mentioned in the algorithm, the power supplied by each generator is modified after each percolation step in order to maintain the power balance in the grid. Figure \ref{fig:sb_of_p} shows the SNBS of each generator node in the complex network as a function of $P$. For both study cases: the Colombian power grid and the randomly grown synthetic power grids, it can be observed that, as the power of the node increases, its stability diminishes; this comes from the fact that $P$ acts as a measure of the stress in the node, that is, the energy demand that it needs to satisfy: having to provide energy for a higher amount of consumer nodes, a generator becomes then more susceptible to random perturbations, as indicated by the reduction on $S_B^{(i)}$. After performing a Pearson correlation testing between $S_B^{(i)}$ and $P^{(i)}$, a correlation coefficient $\sigma$ and a p-value $p_{val}$ was found as $(\sigma, p_{val}) = (-0.605, 0)$ for the Colombian grid and $(\sigma, p_{val}) = (-0.221, 0)$ for the synthetic grids, confirming the existence of a negative relationship between both variables; which is in accordance with the general behavior observed in the recent results shown in \cite{Montanari2020}. This result, as well as the previous observation on the resilience to blackout for $S_B$, can be contrasted with the results presented in \cite{rohden2014NJP}, where it was found that, by distributing the demand among multiple small power generators, instead of a few big power plants, the value of the critical coupling $K_c$ usually diminishes, thus synchronization is favoured, but for disturbances applied to power consumption, the most robust performance was found when there is a mixture of both, small and big power generators.

\subsection{Distribution of vulnerable nodes}

\begin{figure}
\centering
\subfloat[Colombian power grid.]{\includegraphics[width=0.5\linewidth]{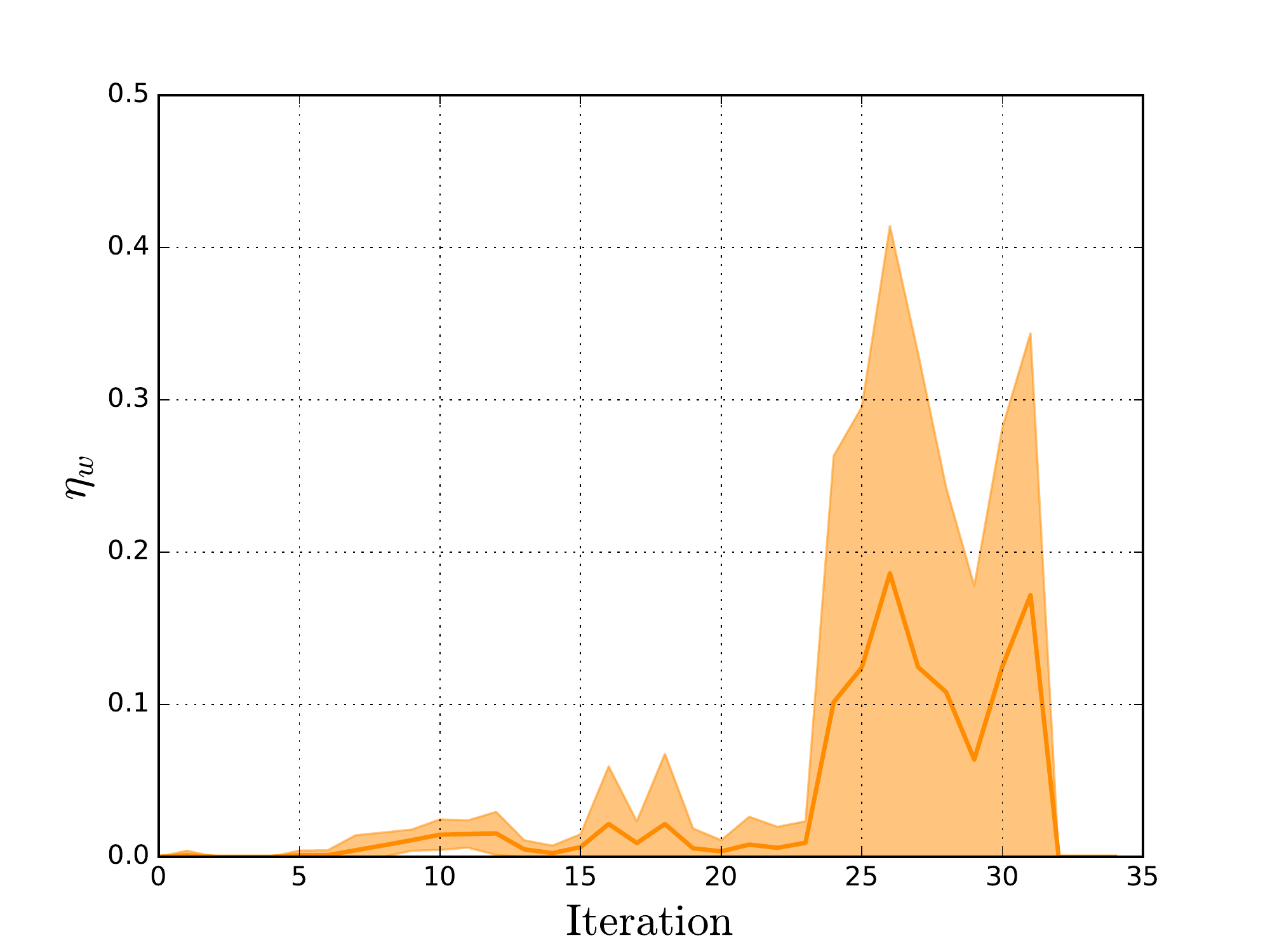}}
\subfloat[Synthetic power grids.]{\includegraphics[width=0.5\linewidth]{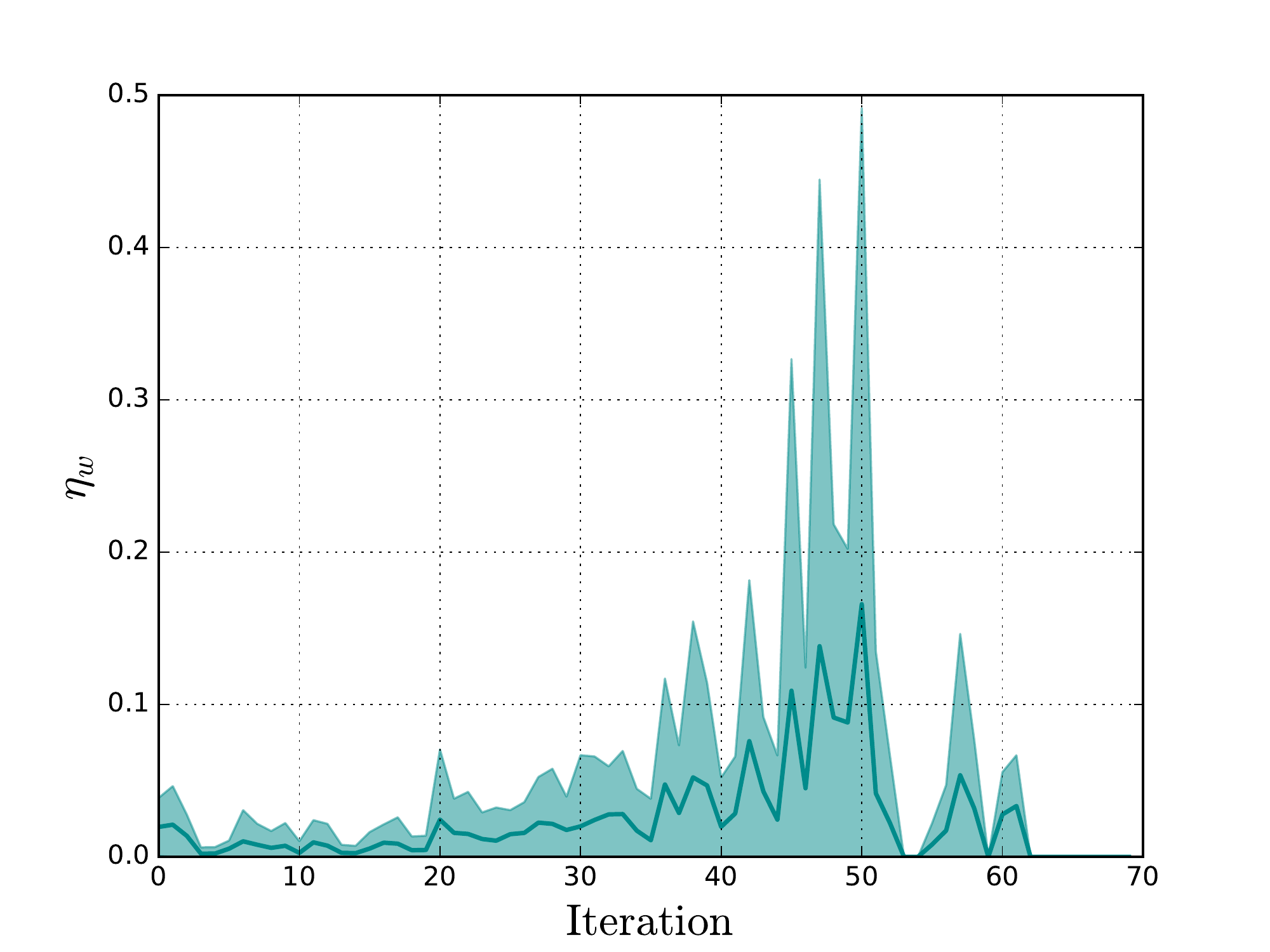}}
\caption{Percentage of weak nodes in the network as a function of the percolation iteration.
\label{fig:n_weak}}
\end{figure}

Let us now consider weak nodes as those that exhibit a SNBS lower than 0.4, then the fraction of weak nodes can be expressed as:
\begin{equation}
\eta_{w(I)} = \frac{N_{w(I)}}{N_{(I)}}
\label{eq:weak_nodes}
\end{equation}
where $N_{w(I)}$ and $N_{(I)}$ are respectively the number of weak nodes and the total amount of nodes in the graph at iteration $I$ of the resilience testing algorithm. Figure \ref{fig:n_weak} shows the behaviour of $\eta_w$ after each iteration for the $S_B$-focused attacking algorithm. From this figure, a rather counterintuitive observation emerges: in general, for both test cases, removing the most vulnerable nodes in the network is causing an overall reduction of the stability of the whole power grid. This implies that, although attacking nodes with a poor SNBS does not ensure a faster transition to blackout than other strategies, it does make the whole grid more prone to dynamical disturbances in oscillation frequency and phase angle. It is worth noting that, results on figure \ref{fig:n_weak} are statistically less reliable in the last iterations due to the fact that each power grid reaches total blackout at a different iteration value in general, thus, last iterations are not averaging necessarily over 10 realizations as the first iterations are.

\begin{figure}
\subfloat[First phase transition.]{\includegraphics[width=0.48\linewidth]{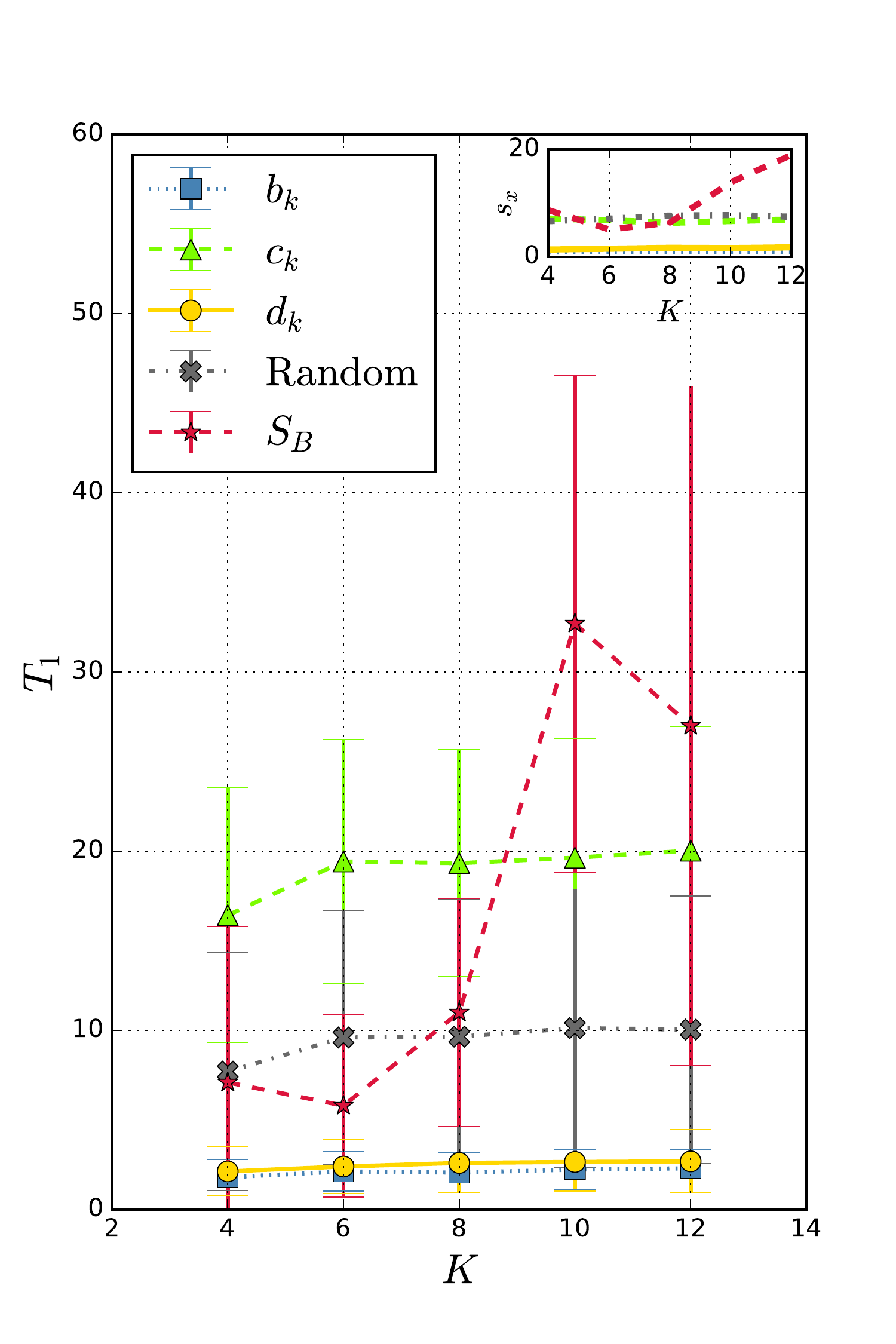}}
\subfloat[Second phase transition.]{\includegraphics[width=0.48\linewidth]{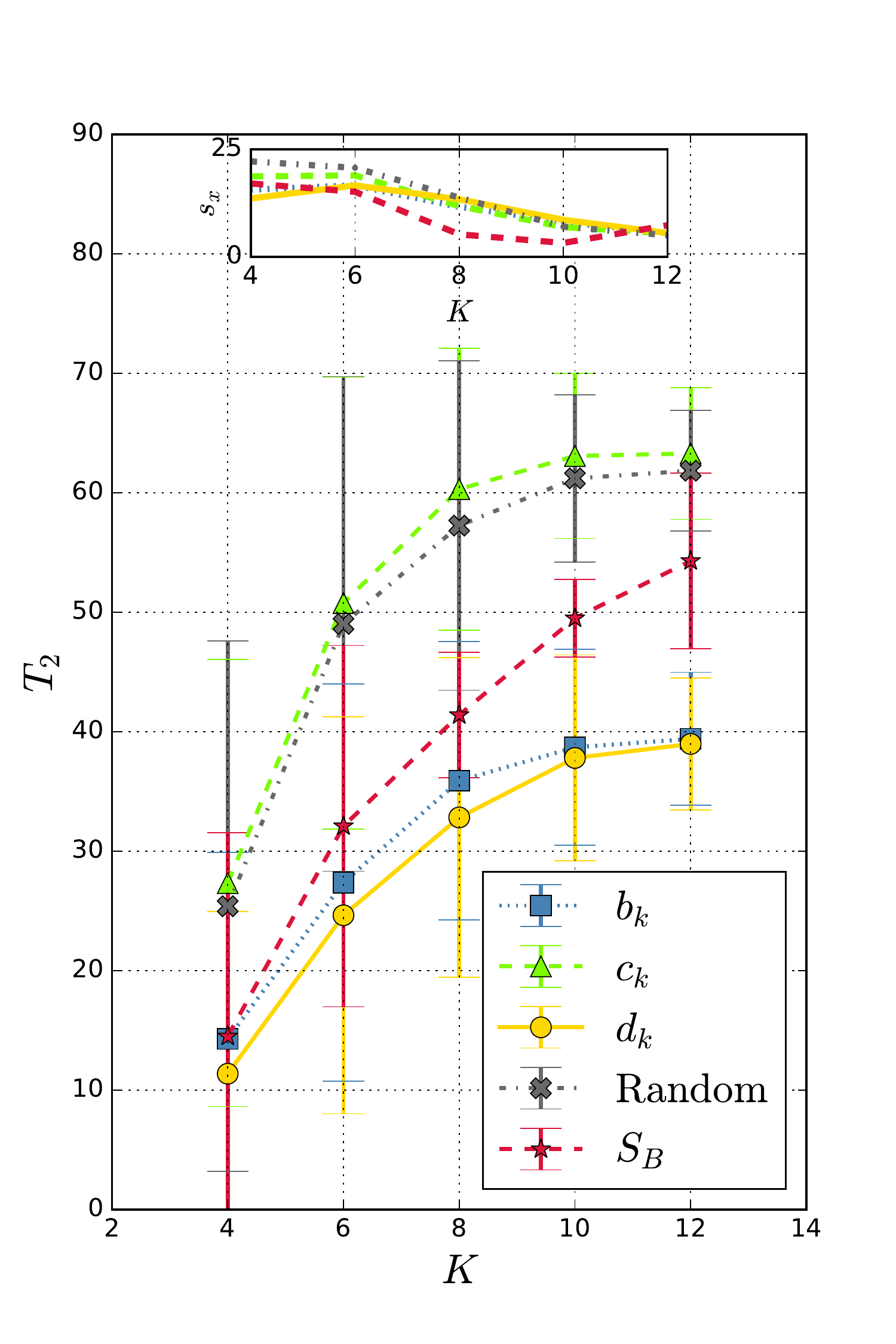}}
\caption{Phase transitions in the node percolation process for synthetic power grids as a function of $K$. Insets present the standard deviation $s_x$ for each computed data point. 
\label{fig:nodetransitions}}
\end{figure}

Another important analysis that can be performed is observing how the critical iterations for which the phase transitions occur, are affected for the parameters in the system. In that regard, for the node resilience procedure over synthetic power grids, $T_1$ and $T_2$ were calculated over variations in the coupling strength $K$ as shown in figure \ref{fig:nodetransitions}, where each data point is the average of 10 simulations for the SNBS method and 500 simulations for the other methods. It should be noted that by decreasing $K$ below 10, $T_1$ for the SNBS case becomes lower than the random case in average, hinting that the effect where weaker nodes are located mainly in inner-tree nodes is more noticeable when $K$ is somewhat large, while for lower $K$, the weak nodes are spread all around the graph. This is understandable since, in general, it can be expected that the basin stability of a node is increased when the coupling of the network is enhanced (this behavior however is not monotonic for all nodes, as pointed out in \cite{Kim2016, Kim2018}). For the random and the topology-dependant elimination methods, no tendency is observed for $T_1$ when varying $K$.

$T_2$ on the other hand shows an evident increase for any removal method when the coupling strength is increased. As shown in figure \ref{fig:nodetransitions}(b), random and topological-based elimination methods seem to reach a plateau when $K$ is sufficiently high. This reduction in the variability of $T_2$ is because for a higher $K$, the network becomes more robust to lose clusters due to lack of synchronization (step 5 in the algorithm) and thus, clusters are removed merely by the lack of generators and consumers (step 3) or by single elimination (step 1 and 2). The number of attacks required to reach blackout on either of those two cases depends mostly on the initial construction of the network, therefore it should not vary significantly among multiple experiments. The SNBS removal depends on the system dynamics, thus its curve does not flat as the others for the observed range, however it is expected to behave in the same way for a significantly higher $K$, when the set point in any node becomes globally stable in $\Pi$.

\begin{figure*}
\subfloat[Giant component of the network.]{\includegraphics[width=0.5\linewidth]{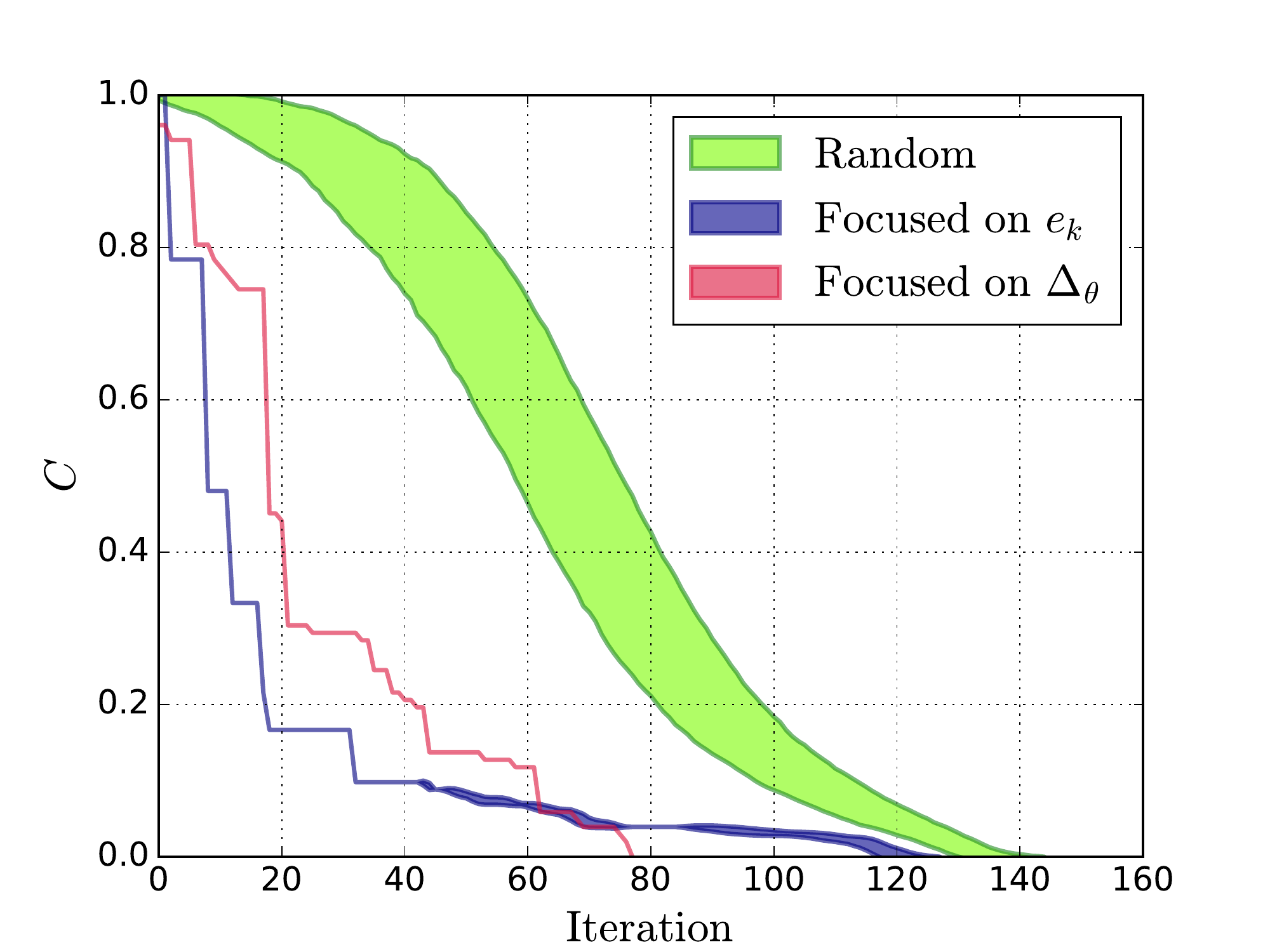}}
\subfloat[Removed component of the network.]{\includegraphics[width=0.5\linewidth]{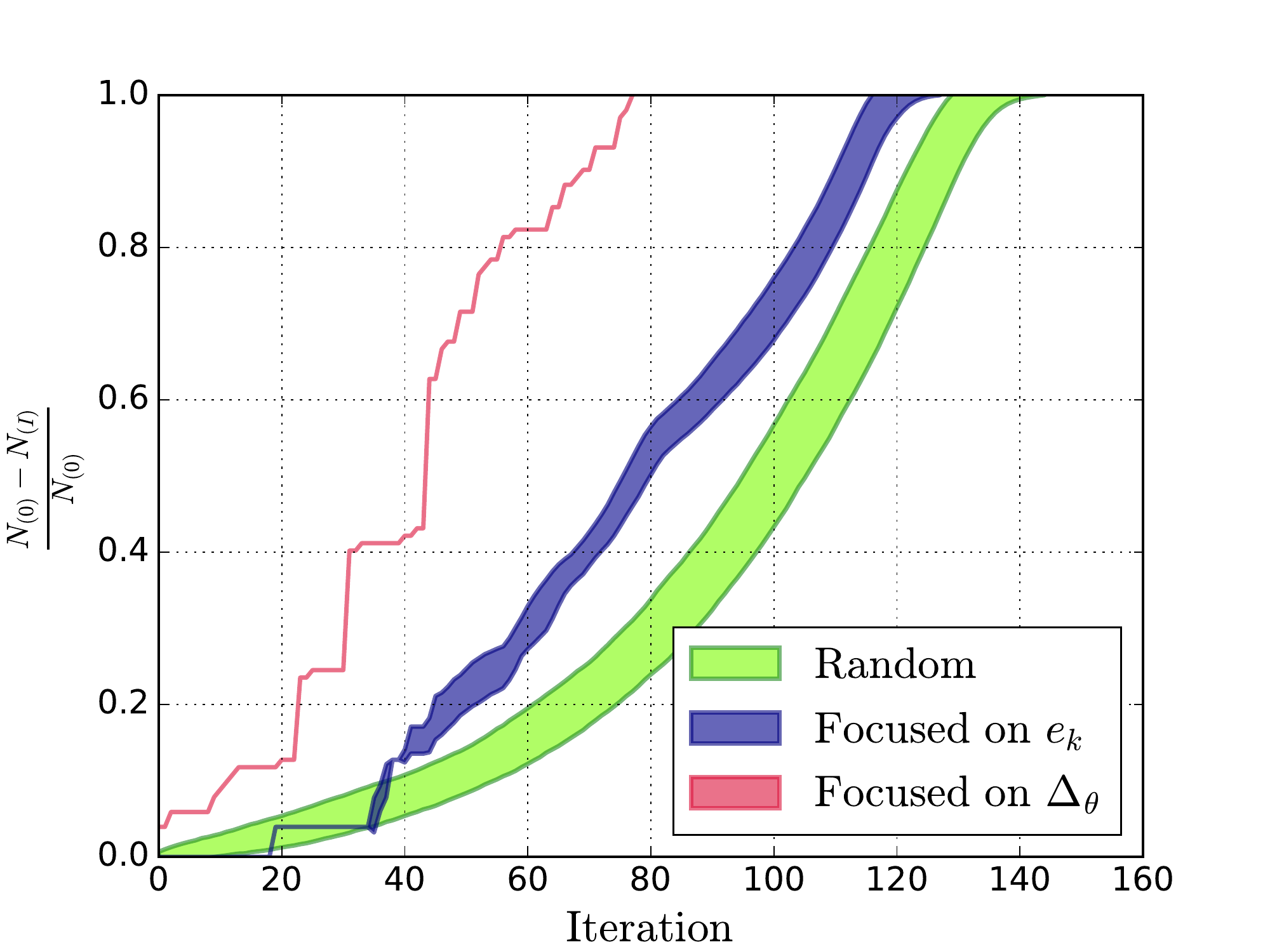}}\\
\subfloat[Second largest component of the network.]{\includegraphics[width=0.5\linewidth]{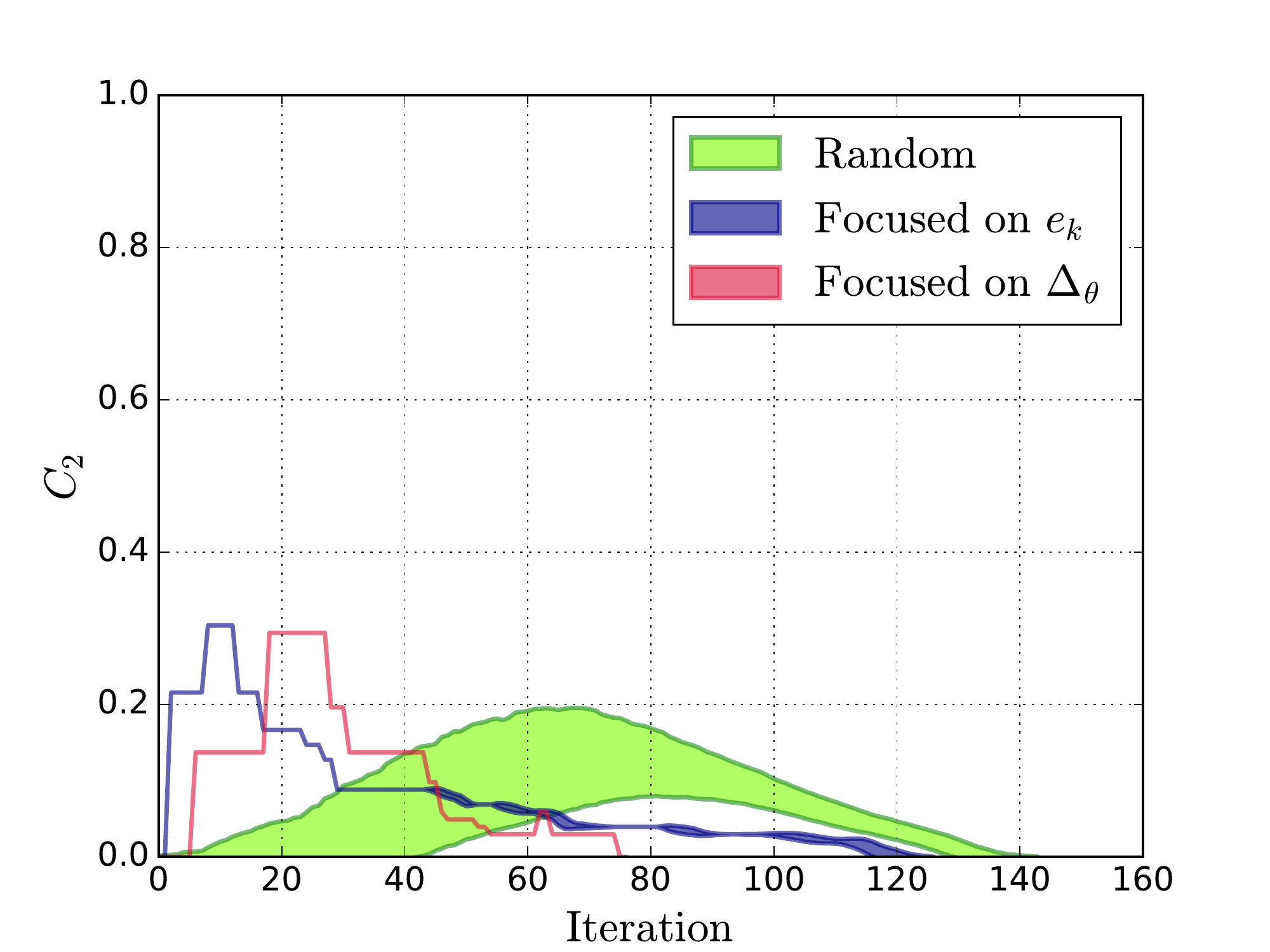}}
\subfloat[Amount of operational clusters in the network.]{\includegraphics[width=0.5\linewidth]{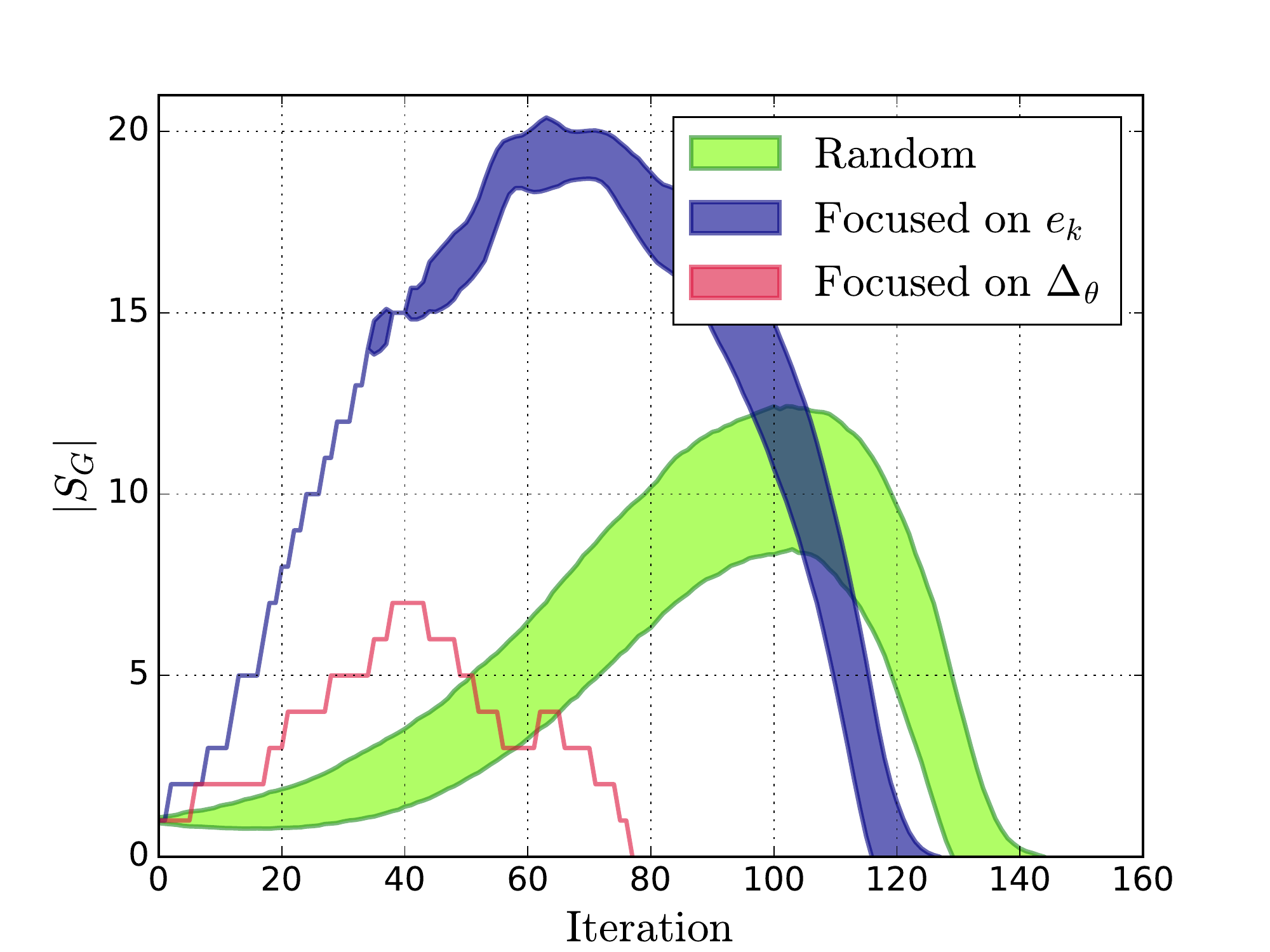}}
\caption{Evolution in the structure of the \textbf{Colombian power grid} subject to the edge elimination algorithm. The shading accounts for the standard deviation calculated over 500 realizations. 
\label{fig:edge_percolation}}
\end{figure*}

\subsection{Edge removal resilience}

Continuing now with the analysis for the edge removal procedure, figure \ref{fig:edge_percolation} presents the evolution of the Colombian graph under this methodology. The difference between the random and focal schemes is remarkable on these simulations: by focusing attacks on transmission lines with higher loads $\Delta_{\theta}(i,j)$, the size of the power grid goes down rapidly and reaches total blackout faster than other alternatives.

\begin{figure*}
\centering
\subfloat[Giant component of the network.]{\includegraphics[width=0.5\linewidth]{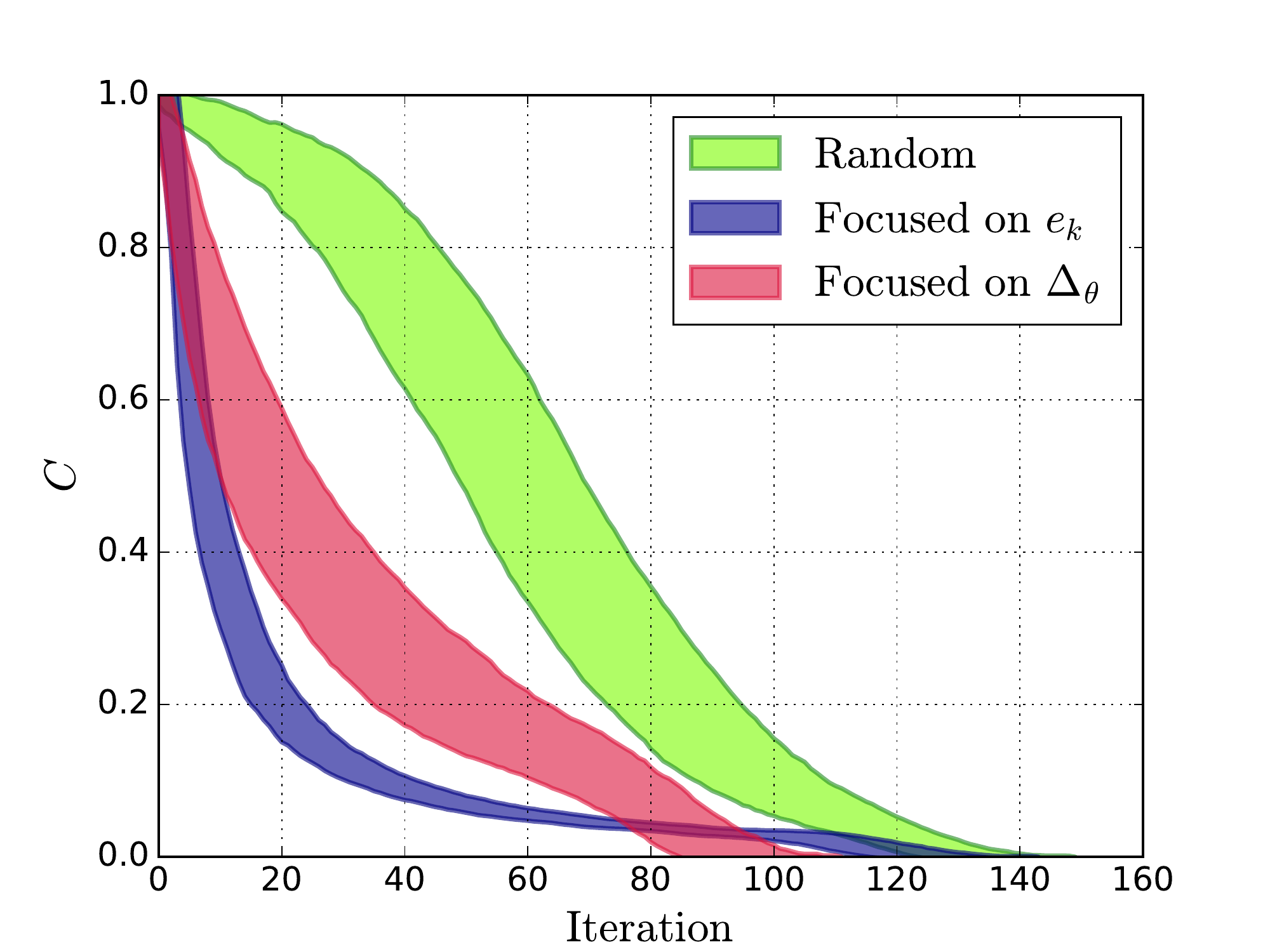}}
\subfloat[Removed component of the network.]{\includegraphics[width=0.5\linewidth]{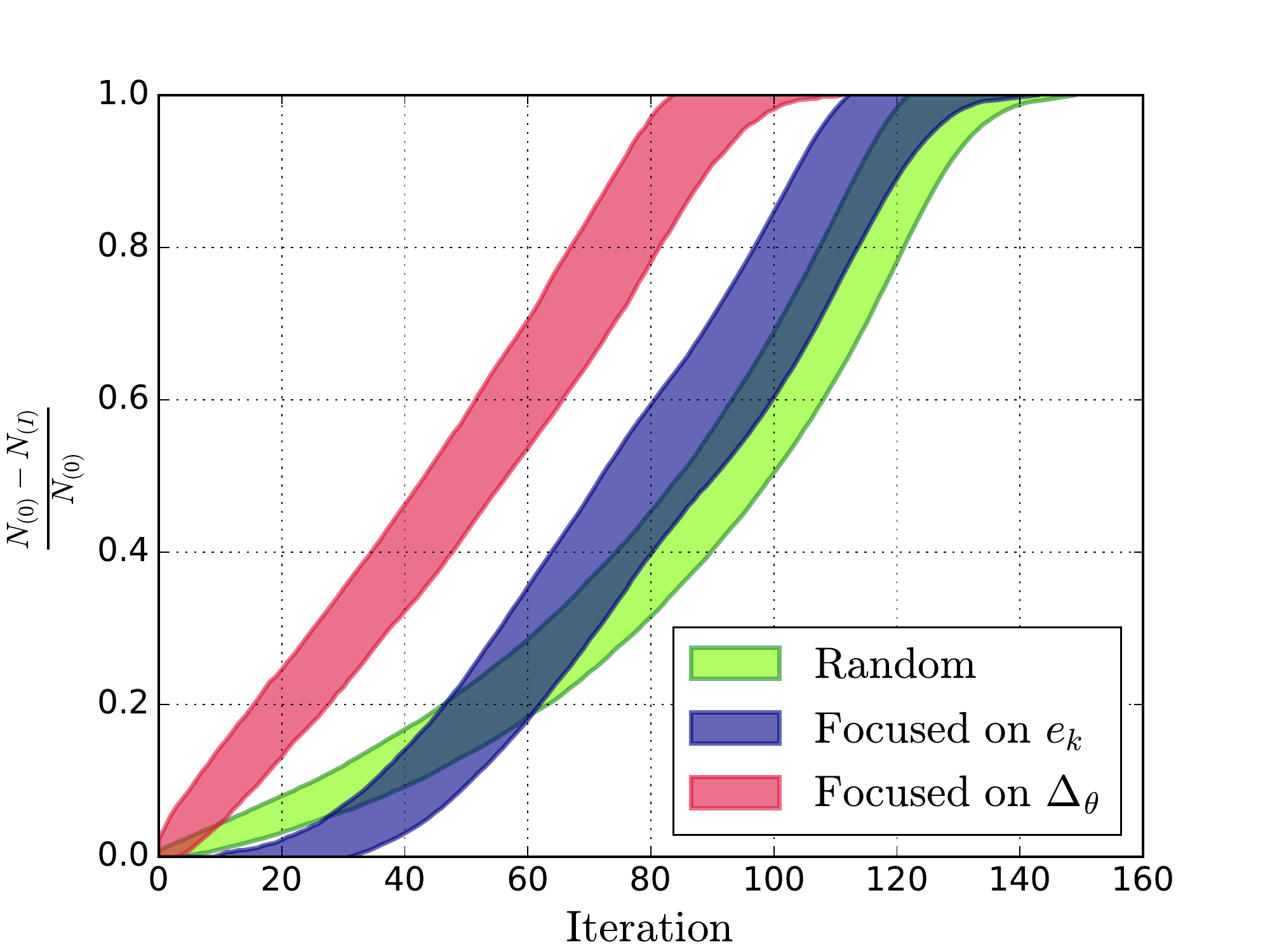}}\\
\subfloat[Second largest component of the network.]{\includegraphics[width=0.5\linewidth]{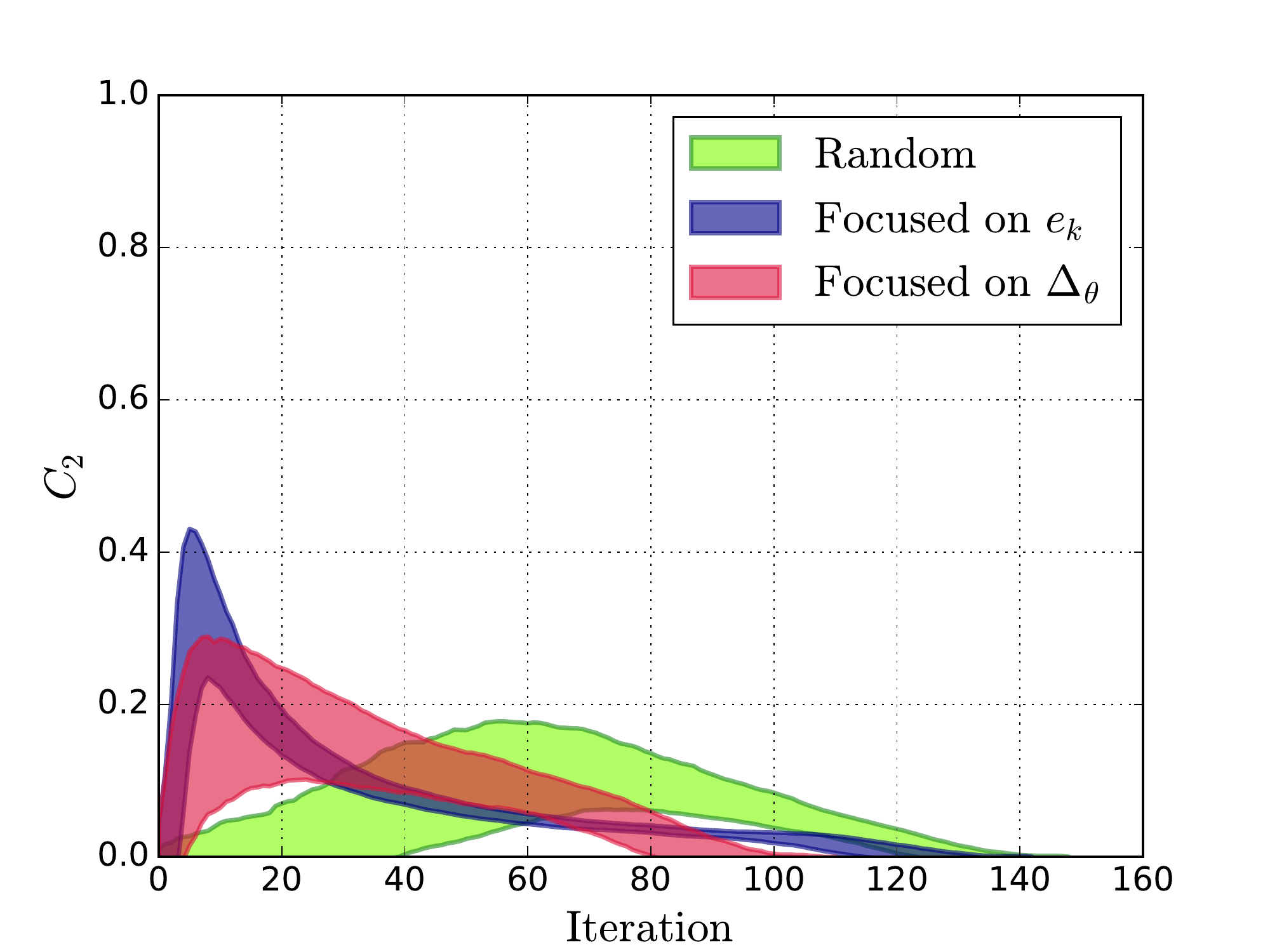}}
\subfloat[Amount of operational clusters in the network.]{\includegraphics[width=0.5\linewidth]{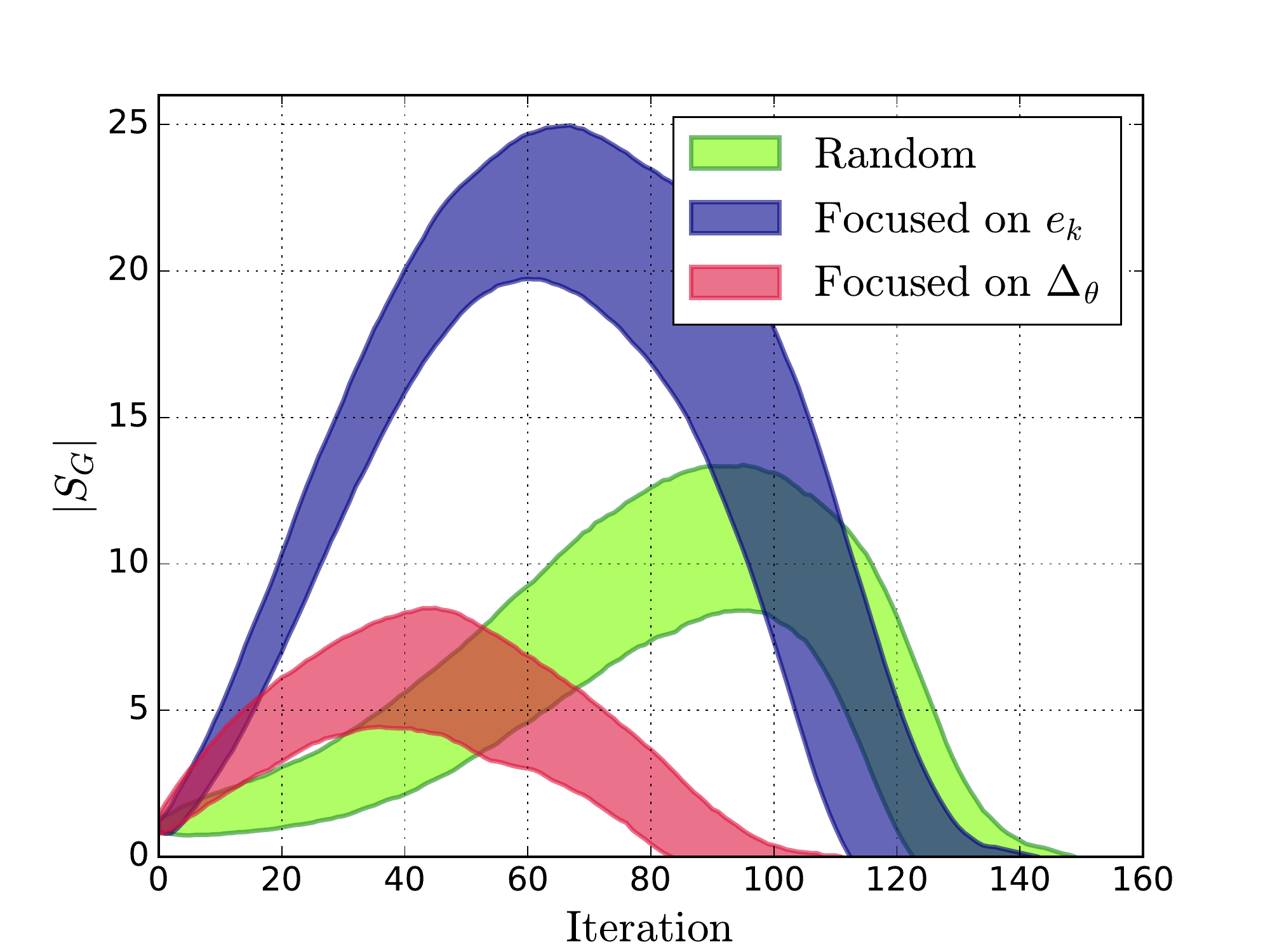}}
\caption{Evolution in the structure of the \textbf{randomly grown synthetic power grids} subject to the edge elimination algorithm. The shading accounts for the standard deviation calculated over 500 realizations. 
\label{fig:edge_percolation_schultz}}
\end{figure*}

By focusing on lines with a higher $e_k$, the sparsity of the graph grows rapidly and many small isolated but functional clusters form, as appreciated in figure \ref{fig:edge_percolation}(d), where the average number of clusters for $e_k$ can even be twice the amount seen in the random case and almost three times the $\Delta_{\theta}(i,j)$ case. Qualitatively, exactly the same behavior is observed on synthetic power grids on figure \ref{fig:edge_percolation_schultz}, and actually, the Colombian grid results lay inside the standard deviation of the experiments for synthetic grids. The values found for $T_1$ and $T_2$ on these simulations are summarized on table \ref{tab:t1t2_edges}.

\begin{table}[h]
\centering
\caption{\label{tab:t1t2_edges}Average value of $T_1$ and $T_2$ for the edge resilience testing algorithm. }
\begin{tabular}{|c|c|c|c||c|c|c|} 
\hline
 & \multicolumn{3}{|c||}{\textbf{Colombian grid}} &
\multicolumn{3}{|c|}{\textbf{Synthetic grids}}\\
\hline
\textbf{} & \textbf{Random} & $\boldsymbol{e_k}$ & $\boldsymbol{\Delta_\theta}$ & \textbf{Random} & $\boldsymbol{e_k}$ & $\boldsymbol{\Delta_\theta}$  \\ 
\hline
$\boldsymbol{T_1}$ & 29.12 & 3.00 & 7.00 & 17.48 & 4.61 & 4.14 \\
\hline
$\boldsymbol{T_2}$ & 131.14 & 117.91 & 77.00 & 125.71 & 116.83 & 87.33\\
\hline
\end{tabular}
\end{table}

\begin{figure}
\subfloat[First phase transition.]{\includegraphics[width=0.48\linewidth]{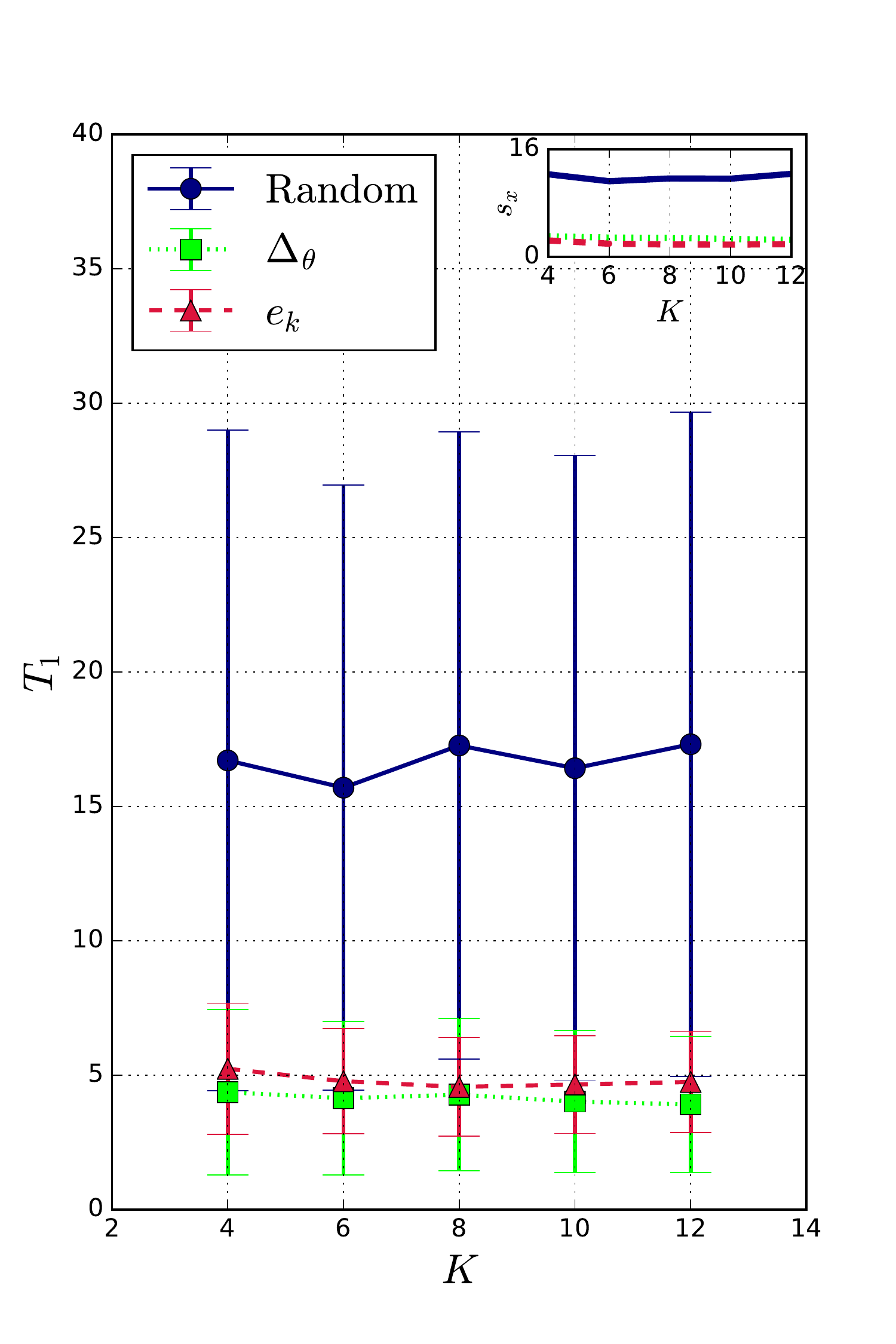}}
\subfloat[Second phase transition.]{\includegraphics[width=0.48\linewidth]{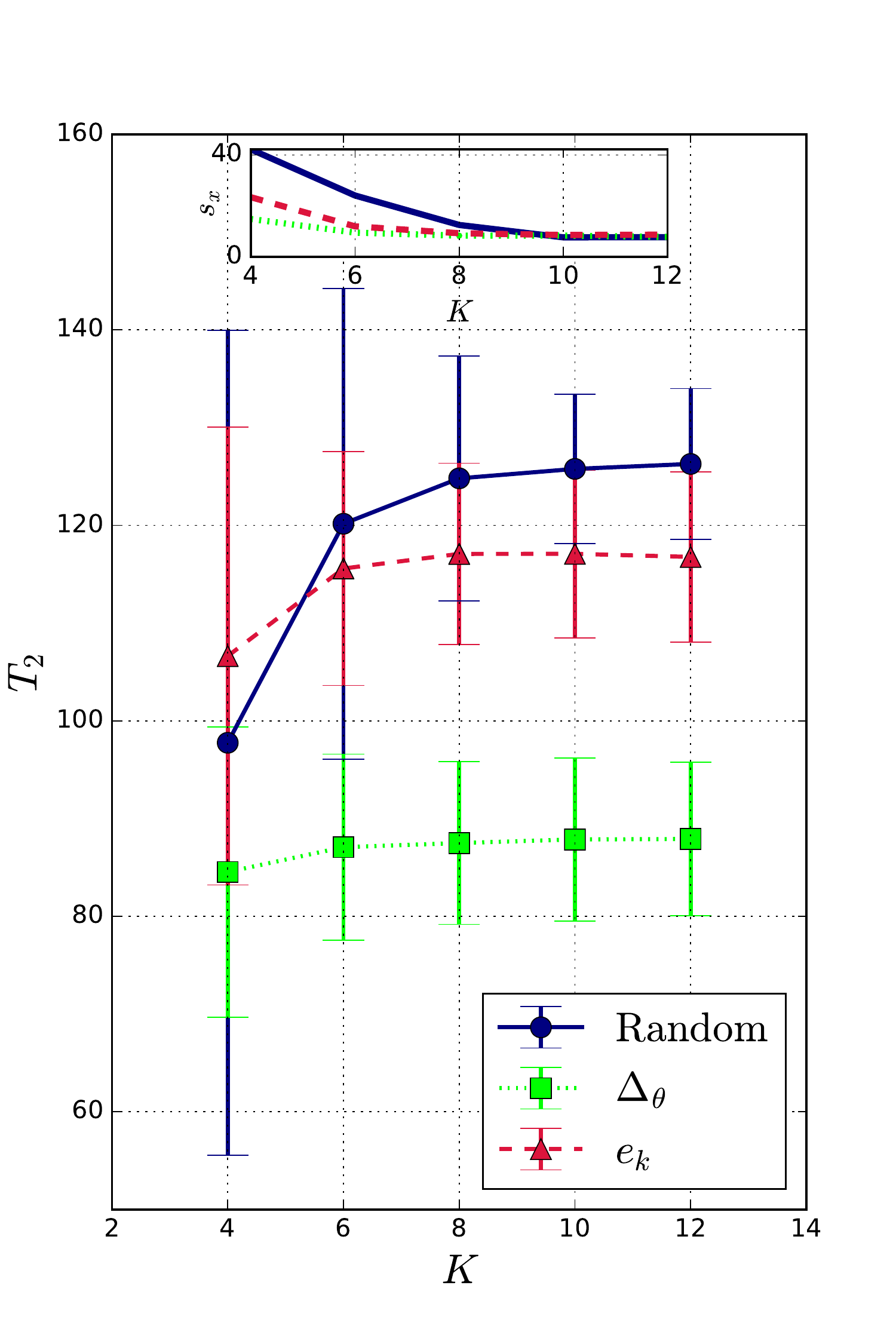}}
\caption{Phase transitions in the edge percolation process for synthetic power grids as a function of $K$. Insets present the standard deviation $s_x$ for each computed data point. 
\label{fig:transitions}}
\end{figure}

Finally, for the line resilience procedure over synthetic power grids, the behavior of the transitions is observed on figure \ref{fig:transitions}. For $T_1$ on panel (a), no clear tendency is observed, besides the expected fact that this transition is, in general, higher for the random method than for other elimination strategies. Panel (b) however shows something interesting for $T_2$: for any value of $K$, attacking the most vulnerable edges based on the dynamics ($\Delta_\theta$) produces the fastest transition to blackout in the synthetic power grids, while attacking randomly produces in general the slowest transition. Furthermore, this transition also reaches the plateau when $K$ is sufficiently large, for any elimination method, as observed for the node attacking results. Evidently, the dynamical-based elimination with $\Delta_\theta$ flats so quickly ($K=6$) because increasing $K$ strengthens the transmission lines and thus directly raises $\Delta_\theta$, as opposed to the node elimination based on SNBS, where the basin stability is known to be affected by $K$ but not in the same trivial and immediate way \cite{Kim2016, Kim2018}.

\section{\label{sec:conclus}Concluding remarks}

In this paper, an algorithm was proposed to test the resilience of power grids against failures. Our approach differs from others found in the literature since it takes into consideration dynamical stability against random finite-size disturbances to describe nodal vulnerability, as well as linear stability of phase differences to compute edge vulnerability. Sequential elimination of elements in the network based on these dynamical vulnerabilities was also compared with the same procedure applied to structural vulnerabilities, which were defined in terms of connectivity patterns in the topology of the graph. This allowed to extract some interesting findings which will be discussed in the following.

First of all, focusing nodal attacks on redundantly connected nodes, such as those with the highest clustering coefficient, is obviously the least effective strategy, being comparable with the random removal; the blackout transition $T_2$ for both is always the highest, meaning that they require more attacks in order to destroy the whole network. Similarly, the giant component of the graph reduces at the slowest rate, since dividing the graph into multiple operational clusters is not likely until the redundant links are removed from the network. Eventually, both methods become essentially the same, once the clustering coefficient for the whole grid is uniform.

The fastest way to reduce the size of the giant component of the network is, as expected, to attack the most central nodes, either those with the highest degree or betweenness; when these bulk nodes are removed, the graph rapidly segregates into multiple perfectly operational clusters. In other words, the transition to an unconnected graph $T_1$ is always the lowest for these methods.

For the case of node elimination based on basin stability it was found that the transition to blackout $T_2$ in the randomly grown synthetic power grids did not differ significantly to that found for the random removal approach. However, in the specific case of the Colombian power grid, this method proved to be the most effective to bring the whole network to total blackout. Essentially, the only key difference between both test cases is the power distribution: for the synthetic grids the power was initially distributed uniformly, with an equal number of generators and consumers, but for the Colombian case only $~27\%$ of the initial nodes are generators, and the rest consumers. This power homogeneity in the synthetic grids then is enhancing the resilience of the graph, and this was confirmed when it was observed that for the Colombian grid, the first nodes to be removed are generators, leading to a higher stress in the remaining power plants: a few generators will have to increase drastically the power they produce in order to balance the demand. In addition, this power increase in the generators will reduce their basin stability, that is, it will make them more prone to dynamical random disturbances in oscillatory frequency and phase angle. This negative correlation between the power parameter of a node and its non-linear stability is supported by the findings in \cite{Montanari2020}. Interestingly, the number of weak nodes in the power grid (nodes with a low basin stability level), was found to increase during the node removal procedure focused on basin stability for both test cases. This implies that this attacking strategy is useful to undermine the general dynamical stability of the power grid and make it more prone to random disturbances, even though, it does not necessarily propagate structural failures to larger proportions of the network (it depends on the amount of generators, as already explained).

Regarding the line resilience testing performed, it was found that for either the Colombian power grid or the synthetic power grids, attacking highly central edges, as those with a high edge betweenness centrality, the graph segregates rapidly into multiple clusters and the giant component reduces abruptly. The transition to blackout, however, is not faster than random elimination. It is, as expected, a very similar behaviour to that found in node elimination through node betweenness. The elimination based on the dynamical measure of phase difference ($\Delta_\theta$) was found to be the most efficient in order to bring the whole network to blackout, and remarkably, this behaviour can be observed for any value tested of the maximum transfer capacity in transmission lines.

Some work is still left open to research such as further exploring the hidden causes that are yielding the reduction of non-linear stability due to heterogeneous power distribution. Similarly, a more detailed study on the phase transitions $T_1$ and $T_2$, as well as the influence of structural parameters in these critical points, could potentially point in the right direction to design the most appropriate connectivity in a graph and parameter distribution such that resilience against cascading failures is optimized. More dynamical measures of vulnerability could also be incorporated in the removal algorithm; for instance, the diffusivity of perturbations \cite{Tamrakar2018} that could indicate regions of faster failure propagation in the grid, or the multi-node basin stability \cite{Mitra2017b} computed over the nearest-neighbours, as it would include information about the most vulnerable clusters, rather than individual elements. This would however impose a higher computational effort to a task that is already computationally expensive.

\section{Acknowledgements}
We would like to pay gratitude to Jorge Fernando Gutiérrez (deceased on september 2019) for valuable ideas and advice 
at the initial stages of this research. We are also thankful with Arthur Montanari, Luis A. Aguirre, Laura Lotero, Fabiola Angulo and
Juan Gabriel Restrepo for fruitful discussions and the technical assistance of Reinel Tabares and Rafael Ruiz. C. C. Galindo-González received financial 
support from \textit{Universidad Nacional de Colombia - Manizales, Dirección de Investigación y Extensión (DIMA)} and 
\textit{Ministerio de Ciencia, Tecnología e Innovación (Minciencias)} under the contracts FP44842 - 052 - 2016 and FP44842 - 505 - 2017. 
D. Angulo-García was financially supported by “Vicerrectoria de Investigaciones - Universidad de Cartagena" through the project 085-2018.

\appendix

\section{\label{random_grids}Random growth model for power grids}

This section describes the algorithm proposed in \cite{Schultz2014} which allows to generate complex networks that incorporate a spatial embedding (accounts for geographic position of nodes) and that have similar structural properties to those found in real-world power grids. For this algorithm, a redundancy cost function that will be subject to optimization has to be defined as:

\begin{equation}
f_{(i,j,G)} = \frac{ ( d_G (i,j) + 1 )^u }{d_S (y_i, y_j)}
\label{eq:optimcost}
\end{equation}

where $d_G (i,j)$ is the length of the shortest path between nodes $i$ and $j$ in the graph $G$, $d_S (y_i, y_j)$ is the spatial distance (Euclidean distance) between the nodes positions $y_i$ and $y_j$, and $u$ is a parameter that controls the cost-vs-redundancy trade-off. After defining the parameters from table \ref{tab:synthetic_grids_params}, the construction algorithm is developed through two phases: initialization and growth, which are performed as follows:

\subsection{Initialization}

\begin{enumerate}
\item Initialize a minimum spanning tree for the initial $N_0$ nodes such that it optimizes the edges that have to be built between them based on the minimum Euclidean distance $d_S (y_i, y_j)$.
\item Let $m = \lfloor N_0 (1 - s)(p + q) \rfloor$. For each $h = \{ 1, 2, ..., m \}$ find the pair of nodes $(i,j)$ that are not yet linked and for which $f_{(i,j,G)}$ is maximal and connect them.\\
\end{enumerate}

\subsection{Growth}

\begin{enumerate}
\item For each $h = \{ 1, 2, ..., N - N_0 \}$ do:

\begin{enumerate}

\item Add a new node $h$ to the graph.
\item Generate a random number $\zeta_1$. If $\zeta_1 \leq 1 - s$ then:

\begin{itemize}
\item[--] Set the position of the new node $y_h$.
\item[--] Add a link between the new node $h$ and the closest node $j$ geographically, that is, the node $j$ for which $d_S (y_h, y_j)$ is minimal.
\item[--] Generate a random number $\zeta_2$, if $\zeta_2 < p$ add a link between the new node $h$ and some node $l$ for which $f_{(h,l,G)}$ is maximal.
\item[--] Generate a random number $\zeta_3$, if $\zeta_3 < q$ draw a node $h'$ from the network uniformly at random. Then find the node $l'$ that is not yet linked to $h'$ and for which$f_{(h,l,G)}$ is maximal and connect them.   
\end{itemize}

\item Otherwise (if $\zeta_1 > 1 - s$), select an edge $(i,j)$ of the network uniformly at random. The new geographic location of node $h$ will be given by $y_h = \nicefrac{(y_i + y_j)}{2}$, so the edge $(i,j)$ is removed from the graph and then the edges $(i,h)$ and $(h,j)$ are added.\\

\end{enumerate}

\end{enumerate}

\section*{References}
\bibliographystyle{iopart-num}
\bibliography{Bibliography_Paper_Power}
\end{document}